\documentclass{pasj00}

\def\labelspace{}
\def\astroph#1{ (astro-ph/#1)}
\def\Umlaut#1{\"{#1}}

\begin{document}
\SetRunningHead{D. Nogami \& T. Iijima}{Spectroscopy of WZ Sge during the 2001 outburst}

\Received{2003/00/00}
\Accepted{2003/00/00}

\title{Dramatic Spectral Evolution of WZ Sagittae during the 2001
Superoutburst}

\author{Daisaku \textsc{Nogami}}
\affil{Hida Observatory, Kyoto University,
       Kamitakara, Gifu 506-1314}
\email{nogami@kwasan.kyoto-u.ac.jp}
\author{Takashi \textsc{Iijima}}
\affil{Astronomical Observatory of Padova, Asiago Section,
       Osservatorio Astrofisico, I-36012 Asiago (Vi), Italy}
\email{iijima@astras.pd.astro.it}



\maketitle

\begin{abstract}

 We carried out optical spectroscopic observations of the most enigmatic
 dwarf nova WZ Sge in 11 nights during the 2001 superoutburst.  Our
 observations covered the period from the initial phase several hours
 before the maximum to the ninth maximum of the rebrightening phase.
 The first spectrum shows absorption lines of H \textsc{i} (except for
 H$\alpha$), He~\textsc{i}, and Na~\textsc{i}, as well as emission lines
 of He\textsc{ii} , C~\textsc{iii}/N~\textsc{iii}, and H$\alpha$ in
 doubly-peaked shapes.  The same spectrum shows the emission lines of
 C~\textsc{iv} and N~\textsc{iv} which are the first detection in dwarf
 novae.  The spectral  features dramatically changed in various time
 scales.  For example, the peak separations of the emission lines of
 H~\textsc{i} and He~\textsc{ii} changed from $\sim$700 km s$^{-1}$ to
 $\sim$1300 km s$^{-1}$, and one of the peaks dominated over an orbital
 period in the genuine-superhump era, but the dominant peak interchanged
 with the orbital phase in the early-superhump era.  The lines of
 H~\textsc{i} and He~\textsc{i} were in emission at minima of the
 rebrightening phase (with no high-excitation lines, nor Na~\textsc{i}),
 while they became in absorption at maxima.  We report on the
 observational results in detail and their implications concerning the
 outburst mechanism, two types of superhumps, and variation of the disk
 structure.

\end{abstract}

\section{Introduction}

WZ Sge is the prototypical star of WZ Sge-type dwarf novae
(\cite{bai79wzsge}; for a recent review, \cite{kat01hvvir}).  This type
of stars is understood as a small group in SU UMa-type dwarf novae, but
has unusual characteristics of quite large outburst amplitudes up to
$\sim$8 mag, extraordinary long recurrence cycles of the outbursts
($\ge$10 years), no (or only a few) normal outbursts (see
\cite{kat01hvvir}; \cite{nog97sxlmi}\footnote{The maximum magnitude of
WZ Sge in table 1 in \cite{nog97sxlmi} should be replaced by $m_o =
8.1$}).  All the outbursts so far observed in WZ Sge have been so-called
superoutbursts in normal SU UMa-type dwarf novae.  Thus the nomenclature
of outburst will be used instead of superoutburst in the case of WZ Sge
stars throughout this paper.

In conjunction with these interesting properties, since WZ Sge is
relatively bright, $m_V \sim 15.3$ in quiescence, many researchers have
most intensively investigated WZ Sge in a variety of methods, and this
star has had a significant effect on many aspects of the study of
cataclysmic variable stars.  While it is impossible to review all the
works related to WZ Sge, we introduce some of most importance.

WZ Sge was first discovered in 1913 as a nova, and the second outburst
in 1946 suggested this star to be a recurrent nova \citep{may46wzsge}.
This second outburst was observed photometrically
(e.g. \cite{him46wzsge}; \cite{ste50wzsge}; \cite{esk63wzsge}), and
spectroscopically \citep{mcl53novaqui}.  \citet{mcl53novaqui} suspected
the dwarf nova nature based on the spectra.  \citet{kra61wzsge} revealed
that WZ Sge is a spectroscopic binary with a period of $\sim$80 min, and
subsequent photometry by \citet{krz62wzsge} proved this star to be an
eclipsing binary star with a period of 81.38 min.

To explain the orbital light curve and the line-profile variation,
\citet{krz64wzsge} for the first time built a model of the cataclysmic
variable star (CVs) containing a white dwarf primary, a late-type
secondary star, and a disk, and deduced binary parameters.  Based on
this model, short-period CVs including WZ Sge were suggested to evolve by
the gravitational wave radiation
\citep{kra62wzsge,pac67CVGWR,fau71GWR,vil71lateevolution,vil71lateevolution2}.
Using the observations of eclipses, \citet{war72wzsge} further developed
this model by including a notion of the hot spot.

High-speed photometry in quiescence by \citet{rob78wzsge} was used to
refine the binary parameters and revealed oscillations with the periods
of 27.87 sec and 28.98 sec.  These oscillations were later interpreted
to be due to the magnetized white dwarf (\cite{pat80wzsge}; see also
\cite{nat78WDoscillation,las99wzsgeIP}).

The third outburst occurred in 1978, which was observed by optical
photometry (e.g. \cite{pat78wzsgeiauc3311,boh79wzsge,hei79wzsge,
bro79wzsge,tar79wzsge,pat81wzsge}), optical spectroscopy
\citep{pat78wzsgeiauc3311,
cra79wzsgespec,gil80wzsgeSH,wal80wzsgespec,ort80wzsge}, and ultraviolet
spectroscopy by the {\it IUE} satellite \citep{fab80wzsgeUV,ort80wzsge,
fri81wzsge}. \citet{pat79SH} suspected the SU UMa nature of WZ Sge,
based on the photometric data of this outburst.  Using the same data,
\citet{pat81wzsge} suggested that the photometric behavior agreed with
that expected by the enhanced mass transfer model, interpreting that
periodic modulations, which are called early superhumps in this paper,
were enhanced orbital humps.  \citet{bai79wzsge} pointed out similarity
of the outburst light curve of UZ Boo and WX Cet with that of WZ Sge,
and proposed that these stars may form a distinct subgroup of the dwarf
novae.

\citet{pap79SHmodel} put forward a new model of an apsidal precession
disk with eccentric orbits to explain the superhumps observed in WZ
Sge and other SU UMa stars.

The orbital period ($P_{\rm orb}$) of WZ Sge had been the shortest one
of the normal hydrogen-rich dwarf novae, although some dwarf novae with
slightly shorter $P_{\rm orb}$ has been found very recently
\citep{tho02gwlibv844herdiuma}.  Its orbital period has put a constraint
on the theory of the CV evolution (for a review of the standard theory,
see e.g. \cite{kin88binaryevolution}).  This is still one of the topics
which are prosperously investigated at present
(\cite{bar03CVminimumperiod}; and references therein).

The small mass of the secondary star in WZ Sge (for recent works,
\cite{ste01wzsgesecondary,ski02wzsgeIR}) has stimulated researchers to
survey brown dwarfs in CVs (e.g. \cite{cia98CVIR, lit00CVIRspec,
pat01SH,men02CVBD, lit03CVBD}.

During the dormancy of WZ Sge after the 1978 outburst, some
outbursts of the members of WZ Sge stars have been observed,
e.g. the 1992 outburst in HV Vir \citep{bar92hvvir,lei93v838her,
kat01hvvir}, the 1995 outburst \citep{pyc95alcom,kat96alcom,how96alcom,
szk96alcomIUE,pat96alcom,nog97alcom} and the 2001 outburst
\citep{ish02wzsgeletter} in AL Com, the 1996--1997 outburst in EG Cnc
\citep{mat98egcnc,liu98egcnc}, the 1998 outburst in V592 Her
\citep{kat02v592cas,men02v592her}, and the 2000--2001 outburst in RZ Leo
\citep{ish01rzleo}.

Among the peculiar properties of WZ Sge stars in outburst these
observations revealed, the definitive behavior commonly seen is the
early superhump clearly distinguishable from the genuine (common)
superhumps [\citet{kat02v592cas} did not find early superhumps
in V592 Cas probably because of lack of observations in the early phase
of the outburst].  The early superhumps have doubly-humped shapes in
contrast to usually singly-humped shape of the genuine superhumps, and
have periods a little, but significantly shorter than the orbital
period (see \cite{ish02wzsgeletter}).  For the early superhumps, some
models have been proposed, e.g. an enhanced hotspot model
\citep{pat81wzsge}, an immature superhump model \citep{kat96alcom}, jet
or axi-asymmetrically flared disk models \citep{nog97alcom}, an
irradiated secondary model\citep{sma93wzsge}.

Theoretical models for the outburst properties mentioned at the top of
this section, the outburst mechanisms, and the rebrightening phenomena
observed in WZ Sge stars have been also published by many researchers,
e.g. \citet{osa95wzsge,war96wzsge,ham97wzsgemodel,mey98wzsge,min98wzsge,
ham99DNevaporation,mey99diskviscosity,ham00DNirradiation,mon01SH,
las01DIDNXT,pat01SH,osa01egcnc,hel01eruma,bua01DNoutburst,
bua02suumamodel}.

As we have reviewed here, WZ Sge is worthy to be called the {\it King of
dwarf novae}.  This dwarf nova gave rise to a new outburst in 2001, 23
years after the previous outburst, and many observation campaigns were
immediately organized.

We here report on the results of our spectroscopic observations
performed during the fourth outburst in 2001.  The next section is
devoted to summarize investigations related to this outburst already
published.  The observations are stated in section 3, and the section 4
describes the results separated into 6 periods.  We reconstruct a story
of the spectral evolution in section 5, and have discussion regarding
the outburst properties, the spectral feature, and the behavior of the
accretion disk in section 6.  Summary and conclusions are put in the
last section 7.

\section{2001 Outburst}

\begin{figure}
 \begin{center}
  \FigureFile(84mm,115mm){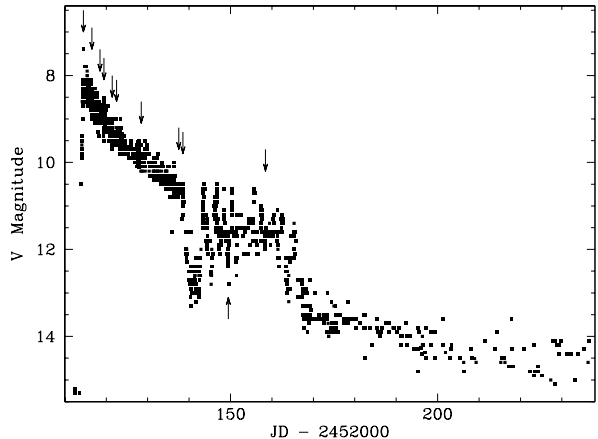}
 \end{center}
 \caption{Long term light curve of the 2001 outburst of WZ Sge.
 The main outburst lasted about 25 days.  After a dip around JD 2452140,
 WZ Sge caused twelve rebrightenings followed by the long fading tail
 (see figure 2 in \cite{ish02wzsgeletter}, or figure 1 in
 \cite{pat02wzsge}).  The dots represent the observations reported to
 VSNET.  The arrows point to the times when our observations were
 carried out.
 }
 \label{fig:lc}
\end{figure}

The 2001 outburst of WZ Sge (figure \ref{fig:lc}) was caught by
T. Ohshima at 2001 July 23.565 (UT) at $m_{\rm vis}=9.7$, which was
first reported to Japanese domestic mailing list by M. Watanabe (vsnet-j
1415)\footnote{$\langle$http://vsnet.kusastro.kyoto-u.ac.jp/vsnet/Mail/j1000/
msg00415.html$\rangle$.}.  R. Ishioka (vsnet-alert
6093)\footnote{$\langle$http://vsnet.kusastro.kyoto-u.ac.jp/vsnet/Mail/alert6000/
msg00093.html$\rangle$.} soon confirmed this outburst and distributed it
via VSNET \citep{VSNET}.  Following photometry on the same night
revealed further brightening to $m_{\rm vis}\sim8.0$ and the presence of
early superhumps \citep{ish01wzsgeiauc7669, mat01wzsgeiauc7669}.
\citet{lan01wzsgeiauc7670} obtained multi-color photometry: $V\sim8.26$,
$U-B\sim-0.97$, $B-V\sim-0.11$, $V-R\sim-0.02$, $R-I\sim-0.03$ about July
25.20 (UT), bluer than the colors in quiescence
(e.g. $B-V=0.10(5), V-R=0.16(5)$ in \cite{mis96sequence}).  Early
superhumps were confirmed by \citet{kat01wzsgeiauc7672}, who measured
the period of the early superhump to be 0.056652(6) d from photometric
data obtained between July 24.46 and 27.01 (see also
\cite{ish02wzsgeletter}).

The spectrum at July 23.74 (UT) had absorption lines of H$\alpha$ and
H$\beta$ superposed on a blue continuum \citep{ish01wzsgeiauc7669}.  In
the next night, however, time-resolved spectra acquired by
\citet{bab01wzsgeiauc7672} between July 24.576 and 24.701 (UT) showed an
emission component in the H$\beta$ absorption line varying with its
orbital period (see \cite{bab02wzsgeletter}).  Doubly-peaked emission
lines of He~\textsc{ii} 4686 and the Bowen blend of
C~\textsc{iii}/N~\textsc{iii} around 4640 \AA\ were also observed in the
same spectra, which were not seen one day before (see figure 1 in
\cite{bab02wzsgeletter}).  From phase-resolved spectra taken between
July 28.02 and 28.24 (UT), \citet{ste01wzsgeiauc7675} drawn Doppler maps
of He~\textsc{ii} 4686 and C~\textsc{iii} 4647 which suggested that the
accretion-disk emission was dominated by two spiral arms (see
\cite{kuu02wzsge}).

The observations by the Chandra X-ray Observatory at July 27.11 (UT)
were characterized by doubly-peaked modulations of the mean 0th-order
count rate with the orbital period and broad (FWHM = 800--1200
km~s$^{-1}$) emission lines of highly ionized species, such as
O~\textsc{v--viii}, Ne~\textsc{v--viii}, Mg~\textsc{v--vii},
Fe~\textsc{vii--ix} \citep{whe01wzsgeiauc7677}.  The spectrum observed
in their second Chandra observation around July 29.73 (UT) was accounted
for with a multi-temperature thermal plasma model with a strong emission
line at 2.4 keV, probably He-like S~\textsc{xv}.

The early superhumps were superseded by the genuine superhumps with a
period of 0.057143(46) d on August 4.53 (UT), 12 days after the start of
the outburst \citep{kat01wzsgeiauc7678}.  The maximum timing of the
superhumps systematically fluctuated around a linear ephemeris [$HJD =
2452126.755 + 0.057153(16)\times E$] with a probable period $\sim$ 4 d,
which were interpreted by the beat between the orbital and the superhump
periods \citep{kat01wzsgeiauc7678}.  \citet{kat01wzsgeiauc7678} also
reported that a doubly-peaked profile of the superhump on August 9 was
changed to a singly-peaked profile on August 11.

WZ Sge rapidly declined in the early days from the maximum, and the
decline then steadily became more gradual.  \citet{can01wzsge} proved
that this trend is a natural result of decrease of the disk mass.  Based
on the growth of the early superhump amplitude around the supermaximum
and the period of the early superhumps 0.05\% longer than $P_{\rm orb}$,
\citet{ish02wzsgeletter} rejected the mass-transfer burst model for the
outburst in WZ Sge originally proposed by \citet{pat81wzsge}.

An asymmetric spiral structure was seen in the Doppler map of the
doubly-peaked emission line He~\textsc{ii} 4686, but not in that of
H$\alpha$, in July 24--27 \citep{bab02wzsgeletter}.  In the emission
line H$\alpha$, \citet{ste01wzsgesecondary} found a narrow emission
component from the irradiated secondary star which first appeared in
early August.  The H$\alpha$ Doppler map constructed by
\citet{ste01wzsgesecondary} from spectra on August 13 indicated that the
disk emission was dominated by a strong extended bright spot.  This map
is much different from those in the early phase of the current outburst
presented by \citet{ste01wzsgeiauc7675} and \citet{bab02wzsgeletter},
but similar to those in quiescence (e.g. \cite{ski00wzsge}).

Quite intensive photometric observations were carried out throughout the
entire outburst with the long fading tail lasting over 100 days, which
are summarized by \citet{pat02wzsge} and \citet{ish04wzsge}.  They
report a variety of photometric behavior in unprecedented detail, such
as early superhumps, common superhumps, development and decay of these
two types of humps, eclipses with a nature different from that in
quiescence, 12 repetitive brightenings following the 3-days dip after
the main outburst, superhumps in the long fading tail.  The limits on
the masses were derived by \citet{pat02wzsge} to be $M_1 > 0.8
M_{\solar}$ and $M_2 < 0.08 M_{\solar}$.

\citet{how03wzsgeIR} obtained time-resolved infrared spectra on 2001
July 27, which contained emission lines of H, He~\textsc{i},
He~\textsc{ii}, Fe~\textsc{i}, Na~\textsc{i}, Ca~\textsc{i},
C~\textsc{i}.  Doppler maps of Pa$\beta$, Pa$\gamma$, He~\textsc{i}, and
He~\textsc{ii} drawn from their spectra showed spiral structures sharing
the same feature with that of He~\textsc{ii} 4686 in optical
\citep{bab02wzsgeletter}.  However, the component with $v_x > 0$
km~s$^{-1}$ was dominant in the maps of Pa$\beta$, Pa$\gamma$, and
He~\textsc{i}.

Far-ultraviolet observations during and after the 2001 outburst
were carried out with the {\it Far Ultraviolet Spectroscopic Explorer}
(FUSE) \citep{lon03wzsgeFUSE} and the {\it Hubble Space Telescope}
(HST) \citep{kni02wzsgeHSToscillation, sio03wzsgeHST}.  The FUSE spectra
on the 7th day of the outburst contained a strong O~\textsc{iv}
absorption line with a blue-shifted core and the absorption lines of
moderate ionization-state ions.  These lines suggest an outflow and a
tenuous layer above the disk (or a vertically extended disk), which is
supported by the HST observations by \citet{sio03wzsgeHST}.
\citet{lon03wzsgeFUSE} also estimated the mass of the white dwarf to be
0.8 $M_{\solar}$, or larger, which is consistent with the results by
\citet{ski00wzsge}, \citet{ste01wzsgesecondary} and \citet{pat02wzsge}.
\citet{kni02wzsgeHSToscillation} observed WZ Sge on 2001 August 8
(plateau phase), 19 (dip), 22 (just after the first peak during the
rebrightening).  Among three data sets, the August 22 one indicated
15-sec oscillations, similar to those seen in quiescence
\citep{pro97wzsge}.  Possible 6.5-sec oscillations on the same night
were first time seen in the history of the WZ Sge study.  Nevertheless,
these oscillations were not caught on August 8 and 19.
\citet{kni02wzsgeHSToscillation} did not find evidence of 29-sec signals
which had been detected by \citet{rob78wzsge}, \citet{pat80wzsge},
\citet{ski97wzsge}, \citet{wel97wzsgeUVoscillation}, \citet{pat98wzsge},
\citet{ski99wzsge}, and \citet{ski02wzsgeIR}.

Based on the detailed observations of the early superhumps presented by
\citet{ish02wzsgeletter} and \citet{pat02wzsge}, two new models have
been proposed by \citet{osa02wzsgehump} and \citet{kat02wzsgeESH}.
\citet{osa02wzsgehump} described that early superhumps can be explained
by tidal removal of the angular momentum from the accretion disk by the
2:1 resonance \citep{lin79lowqdisk}\footnote{Interestingly,
\citet{lin79lowqdisk} originally introduced the idea of a spiral
dissipation pattern due to the 2:1 resonance to explain the
doubly-peaked shape of the orbital humps of WZ Sge observed in
quiescence.}.  The model by \citet{kat02wzsgeESH} is an application of
the tidal distortion effect in the accretion disk (see
\cite{sma02ADstructure, ogi02tidal}).

\begin{table*}
\caption{Log of the observations.}
\begin{center}
\begin{tabular}{lrccccrcccc}
\hline\hline
\multicolumn{2}{c}{Date} & Start & Exposure & File ID & Mid BDJD  & Orbital &
Instr. & Spectral & Period & Comments \\
 & & UT & Time(s) & & (2400000+) & Phase  & & Range (\AA) & & \\ \hline
Jul.& 23 & 23:18 &  300 & 10550 & 52114.47802 &   0.255 &B\&C & 3942--5130 & I & \\
    &    & 23:41 &  300 & 10553 & 52114.49368 &   0.531 &  "  & 4844--6041 & " & \\
    & 24 & 00:08 &   60 & 10557 & 52114.51157 &   0.847 &  "  & 5698--6901 & " & \\
    &    & 00:10 &  300 & 10558 & 52114.51437 &   0.896 &  "  &      "	    & " & \\
    & 25 & 22:49 &  300 & 10562 & 52116.45794 &  35.182 &  "  & 3984--5173 & " & cloudy \\
    & 26 & 00:03 &  300 & 10564 & 52116.50894 &  36.081 &  "  &      "	    & " & \\
    &    & 00:40 &  300 & 10568 & 52116.53492 &  36.540 &  "  & 4767--5963 & " & \\
    &    & 02:11 &  180 & 10577 & 52116.59774 &  37.648 &  "  & 5818--7021 & " & \\
    &    & 02:15 &  300 & 10578 & 52116.60102 &  37.706 &  "  &     "	    & " & \\
    &    & 02:25 &  600 & 10580 & 52116.60967 &  37.858 &  "  & 3983--5173 & " & \\
    & 27 & 21:30 &  600 & 10585 & 52118.40481 &  69.525 &  "  & 3985--5174 & " & \\
    &    & 21:57 &  600 & 10588 & 52118.42319 &  69.850 &  "  &      "     & " & \\
    &    & 23:39 &  600 & 10595 & 52118.49438 &  71.106 &  "  &      "     & " & cloudy \\
    &    & 23:54 &  300 & 10597 & 52118.50300 &  71.258 &  "  & 5771--6974 & " & \\
    & 28 & 01:43 &  300 & 10603 & 52118.57875 &  72.594 &  "  &      "     & " & \\
    &    & 22:48 &  600 & 10613 & 52119.45913 &  88.124 &  "  & 3937--5125 & II &  cloudy \\
    & 30 & 21:38 &  360 & 10616 & 52121.40921 & 122.525 &  "  & 3924--5111 & " & \\
    &    & 21:48 &  600 & 10618 & 52121.41743 & 122.670 &  "  &      "     & " & \\
    &    & 22:20 &  360 & 10620 & 52121.43779 & 123.029 &  "  &      "     & " & \\
    &    & 22:28 &  360 & 10622 & 52121.44388 & 123.136 &  "  &      "     & " & \\
    &    & 22:37 &  360 & 10624 & 52121.44961 & 123.237 &  "  &      "     & " & \\
    &    & 22:43 &  360 & 10625 & 52121.45427 & 123.319 &  "  &      "     & " & \\
    &    & 22:53 &  360 & 10626 & 52121.46127 & 123.443 &  "  &      "     & " & \\
    &    & 23:02 &  360 & 10628 & 52121.46729 & 123.549 &  "  &      "     & " & \\
    &    & 23:09 &  360 & 10629 & 52121.47199 & 123.632 &  "  &      "     & " & \\
    &    & 23:16 &  360 & 10630 & 52121.47708 & 123.722 &  "  &      "     & " & \\
    &    & 23:23 &  360 & 10631 & 52121.48152 & 123.800 &  "  &      "     & " & \\
    & 31 & 21:43 &  600 & 10637 & 52122.41396 & 140.249 &  "  & 4038--5227 & " & \\
    &    & 21:56 &  600 & 10639 & 52122.42288 & 140.406 &  "  &      "     & " & cloudy \\
    &    & 22:09 &  600 & 10640 & 52122.43203 & 140.568 &  "  &      "     & " & \\
    &    & 22:21 &  600 & 10642 & 52122.44016 & 140.711 &  "  &      "     & " & \\
    &    & 22:33 &  600 & 10644 & 52122.44844 & 140.857 &  "  &      "     & " & \\
    &    & 22:43 &  600 & 10645 & 52122.45578 & 140.987 &  "  &      "     & " & \\
    &    & 22:54 &  600 & 10646 & 52122.46309 & 141.116 &  "  &      "     & " & \\
    &    & 23:06 &  600 & 10648 & 52122.47153 & 141.264 &  "  &      "     & " & \\
    &    & 23:17 &  600 & 10649 & 52122.47887 & 141.394 &  "  &      "     & " & \\
    &    & 23:27 &  600 & 10650 & 52122.48619 & 141.523 &  "  &      "     & " & \\
Aug.&  1 & 00:00 &  300 & 10654 & 52122.50691 & 141.889 &  "  & 5775--6979 & " & \\
    &    & 00:07 &  360 & 10656 & 52122.51234 & 141.984 &  "  &      "     & " & \\
    &    & 00:13 &  360 & 10657 & 52122.51670 & 142.061 &  "  &      "     & " & \\
    &    & 00:20 &  360 & 10658 & 52122.52125 & 142.142 &  "  &      "     & " & \\
    &    & 00:30 &  360 & 10659 & 52122.52807 & 142.262 &  "  &      "     & " & \\
    &    & 00:39 &  360 & 10662 & 52122.53460 & 142.377 &  "  &      "     & " & \\
    &    & 00:46 &  360 & 10663 & 52122.53933 & 142.460 &  "  &      "     & " & \\
    &    & 00:52 &  360 & 10664 & 52122.54398 & 142.543 &  "  &      "     & " & \\
    &    & 01:00 &  360 & 10665 & 52122.54942 & 142.638 &  "  &      "     & " & \\
    &    & 01:09 &  360 & 10667 & 52122.55512 & 142.739 &  "  &      "     & " & \\
    &    & 01:16 &  360 & 10668 & 52122.56011 & 142.827 &  "  &      "     & " & \\
    &    & 01:23 &  360 & 10669 & 52122.56498 & 142.913 &  "  &      "     & " & \\
    &    & 01:30 &  360 & 10670 & 52122.57006 & 143.002 &  "  &      "     & " & \\
    &    & 01:38 &  360 & 10672 & 52122.57538 & 143.096 &  "  &      "     & " & \\
    &    & 01:45 &  360 & 10673 & 52122.58052 & 143.187 &  "  &      "     & " & \\
    &    & 01:52 &  360 & 10674 & 52122.58564 & 143.277 &  "  &      "     & " & \\
    &    & 02:00 &  360 & 10675 & 52122.59058 & 143.364 &  "  &      "     & " & \\
    &    & 02:07 &  360 & 10676 & 52122.59541 & 143.450 &  "  &      "     & " & \\ 
    &    & 02:15 &  360 & 10678 & 52122.60121 & 143.552 &  "  &      "     & " & cloudy \\
    &    & 02:22 &  360 & 10679 & 52122.60602 & 143.637 &  "  &      "     & " & cloudy \\
\hline
\end{tabular}
\end{center}
\end{table*}

\addtocounter{table}{-1}
\begin{table*}
\caption{(continued)}\label{tab:log}
\begin{center}
\begin{tabular}{lrccccccccc}
\hline\hline
\multicolumn{2}{c}{Date} & Start & Exposure & File ID & Mid BDJD & Orbital  & Instr. &
Spectral & Period & Comments \\
 & & UT & Time (s) & &  (2400000+) & Phase &        &
Range (\AA) & & \\ \hline
Aug.&  6 & 22:09 &  900 & 37106 & 52128.43328 & 246.432  &Echelle& 4327--6891 & III & \\
    &    & 22:28 &  900 & 37107 & 52128.44668 & 246.669  &  "  &      "       & " & \\
    &    & 22:46 &  600 & 37109 & 52128.45773 & 246.864  &  "  &      "       & " & \\
    &    & 22:58 &  600 & 37110 & 52128.46569 & 247.004  &  "  &      "       & " & \\
    &    & 23:09 &  600 & 37111 & 52128.47366 & 247.145  &  "  &      "       & " & \\
    &    & 23:21 &  600 & 37112 & 52128.48182 & 247.289  &  "  &      "       & " & \\
    &    & 23:35 &  600 & 37114 & 52128.49170 & 247.463  &  "  &      "       & " & \\
    &    & 23:47 &  600 & 37115 & 52128.49971 & 247.604  &  "  &      "       & " & \\
    &    & 23:58 &  600 & 37116 & 52128.50762 & 247.744  &  "  &      "       & " & \\
    &  7 & 00:10 &  600 & 37117 & 52128.51557 & 247.884  &  "  &      "       & " & \\
    &    & 00:21 &  600 & 37118 & 52128.52348 & 248.024  &  "  &      "       & " & \\
    &    & 00:45 &  600 & 37121 & 52128.53984 & 248.312  &  "  &      "       & " & \\
    &    & 00:56 &  600 & 37122 & 52128.54783 & 248.453  &  "  &      "       & " & \\
    & 16 & 00:33 &  600 & 10729 & 52137.53196 & 406.937  &B\&C & 3999--5188   & IV & \\
    &    & 00:44 & 1200 & 10730 & 52137.54310 & 407.134  &  "  &      "       & " & \\
    &    & 01:07 & 1200 & 10732 & 52137.55883 & 407.411  &  "  &      "       & " & \\
    &    & 23:26 & 1200 & 10740 & 52138.48867 & 423.814  &  "  & 3998--5187   & " & \\
    &    & 23:48 & 1200 & 10742 & 52138.50420 & 424.088  &  "  &      "       & " & \\
    & 17 & 00:11 & 1200 & 10744 & 52138.52008 & 424.368  &  "  &      "       & " & \\
    &    & 00:35 & 1200 & 10746 & 52138.53647 & 424.657  &  "  &      "       & " & \\
    & 27 & 22:19 & 1800 & 10993 & 52149.44533 & 617.095  &  "  & 3948--5136   & V & \\
    &    & 23:00 & 1800 & 10995 & 52149.47395 & 617.600  &  "  &      "       & " & \\
    &    & 23:37 & 1800 & 10997 & 52149.49919 & 618.045  &  "  &      "       & " & \\
    & 28 & 00:53 & 1200 & 11003 & 52149.54910 & 618.925  &  "  & 5772--6976   & " & \\
Sep.&  5 & 22:16 & 2400 & 11021 & 52158.44629 & 775.876  &  "  & 3938--5125   & VI & \\
\hline
\end{tabular}
\end{center}
\end{table*}

\citet{osa03DNoutburst} critically examined the ``observed'' evidence of
enhanced mass transfer including that proposed by \citet{pat02wzsge},
and argued that the overall light curve of the 2001 outburst does
not require an assumption of enhanced mass transfer in the scheme of the
thermal-tidal disk instability model.

Being inspired by the complex superhump light curves observed during the
2001 outburst in WZ Sge, \citet{osa03wzsgetomography} proposed a
new method ``helical tomography'', to analyze superhump light curves of
SU UMa stars with high orbital inclination.

\section{Observation}

We obtained 83 optical spectra in total in 11 nights during the 2001
outburst (figure \ref{fig:lc}).  Table \ref{tab:log} gives a
journal of the observations.  The spectra of a medium resolution
($\sim 6$ \AA) were taken with a Boller \& Chivens spectrograph mounted
on the 122-cm telescope of the Asiago Astrophysical Observatory of the
University of Padova, using a 512$\times$512-pixel CCD detector.  An
echelle spectrograph on the 182-cm telescope of Mount Ekar station of
the Astronomical Observatory of Padova and a CCD camera of
550$\times$550 pixels were used to obtain the high-dispersion
spectroscopy (resolution 0.6 \AA).

The spectra were reduced in the standard ways using the IRAF
package\footnote{IRAF is distributed by the National Optical Astronomy
Observatories for Research in Astronomy, Inc. under cooperative
agreement with the National Science Foundation.} at the Asiago
Observatory.  The sensitivity of the spectrographs were corrected using
spectra of some spectrophotometric standard stars (BD+25~3941 and
HD~192281 for the medium dispersion spectroscopy, and 58 Aql for the
high dispersion spectroscopy) obtained in the same nights.  The
signal-to-noise ratio at the continuum varies over 100 to $\sim$20,
depending on the sky condition, the exposure time, and, most
significantly, the object brightness.  The orbital phase $\phi$, which
is used throughout this paper, is calculated with the following
ephemeris:
\begin{equation}
\phi = \frac{{\rm BDJD} - T_0}{P_{\rm orb}} - \phi_{\rm cor} - E_0,
\end{equation}
where $T_0$ = 2437547.728868, $P_{\rm orb}$ = 0.05668784707, and
$\phi_{\rm cor}$ = $-$0.022 (see \cite{ski00wzsge}).  The $E_0$ is set
to be 256964 in order that the orbital phase in this paper starts around
$\phi=0$.

\section{Results}

In this section, we describe detailed spectral feature, separating our
observations into six periods.  As in table \ref{tab:log}, period I is
the very early phase of the outburst from JD 2452114 to 2452118,
period II is from 2452119 to 2452122, period III is on JD 2452128,
period IV is the end of the main outburst on JD 2452137 and 2452138,
period V is in the third minimum in the rebrightening stage on JD
2452149, and period VI is at the 9th rebrightening peak on JD
2452158.

\begin{figure}
  \begin{center}
    \FigureFile(84mm,115mm){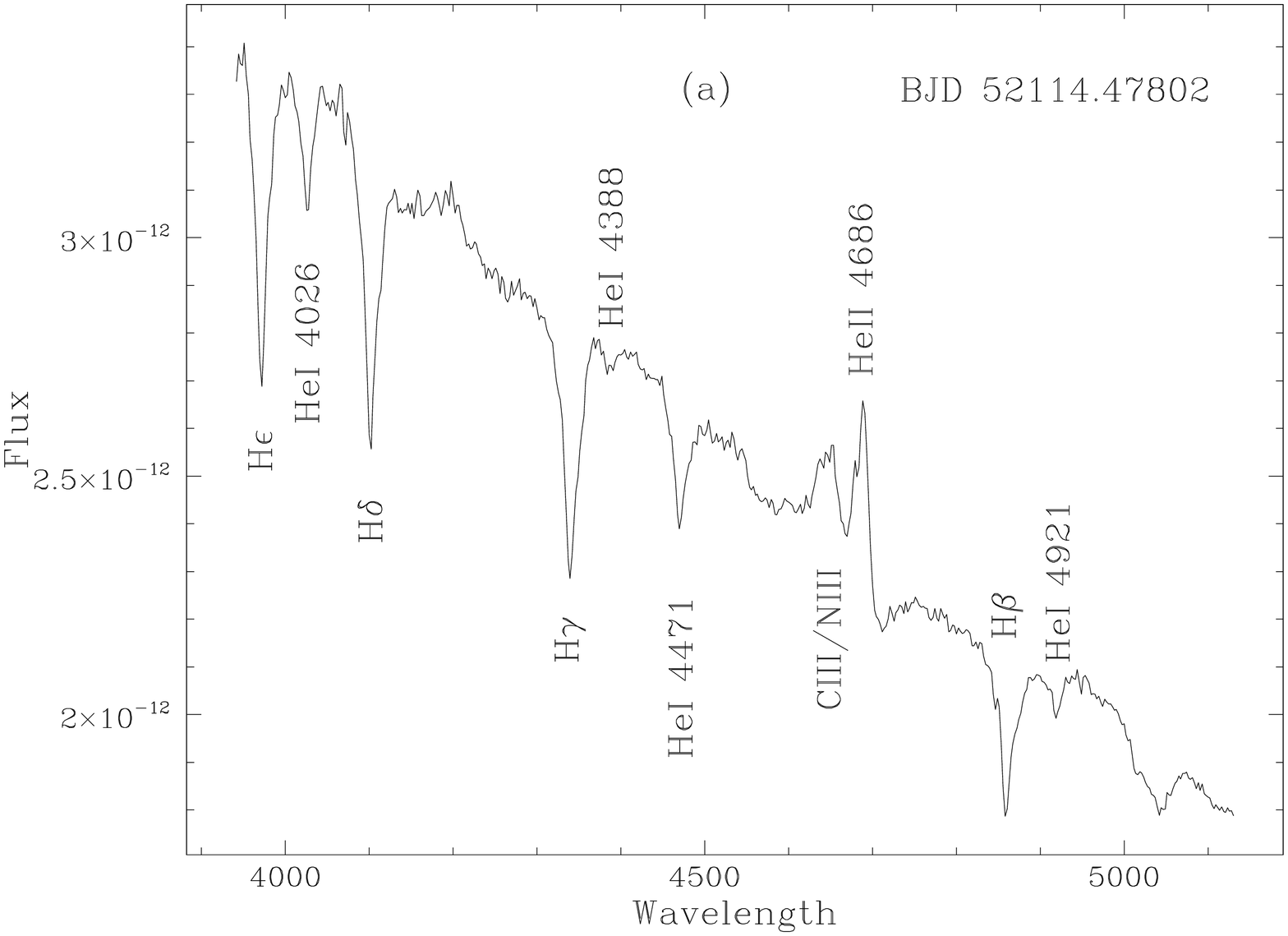}\\
    \FigureFile(84mm,115mm){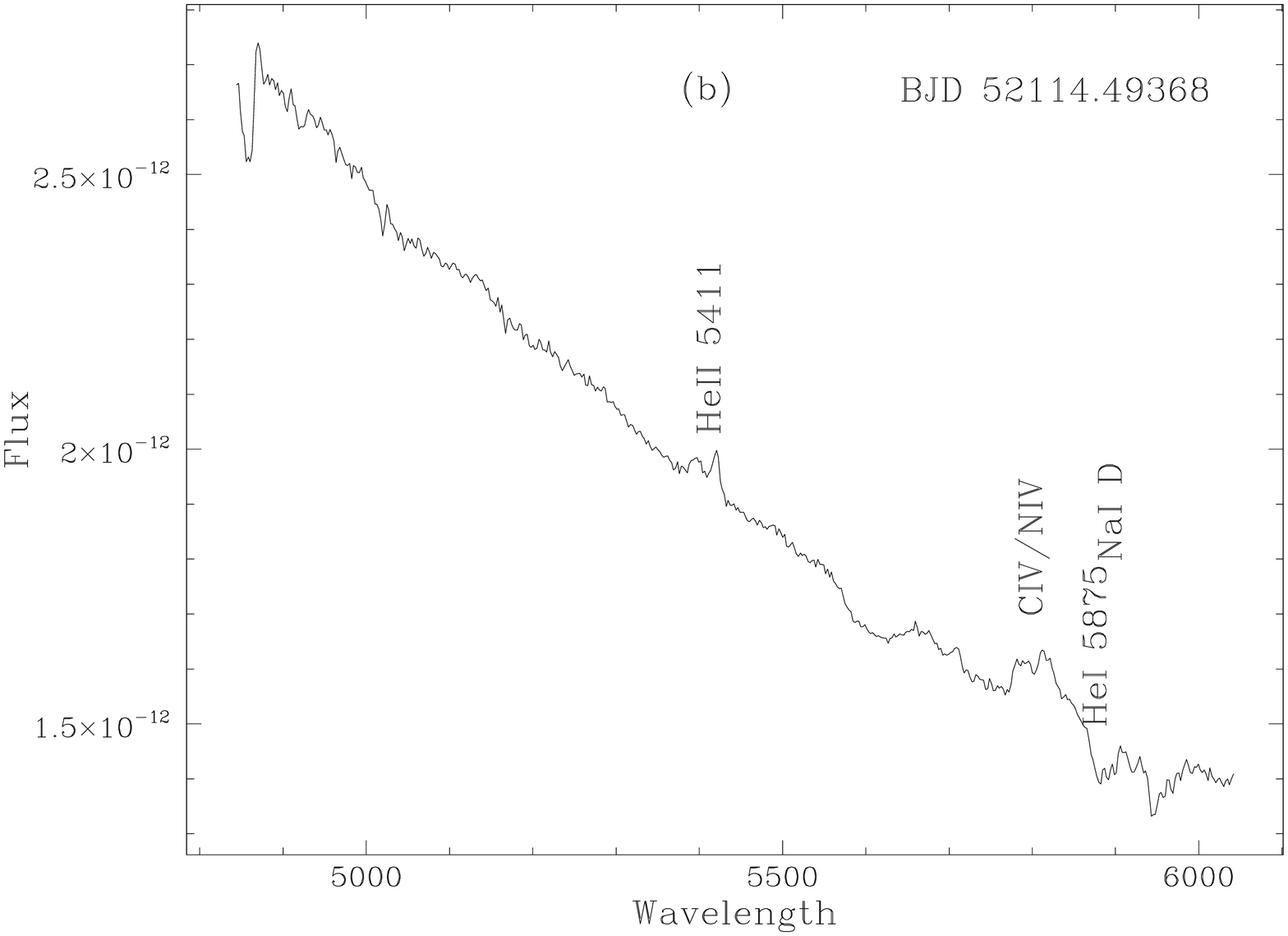}\\
    \FigureFile(84mm,115mm){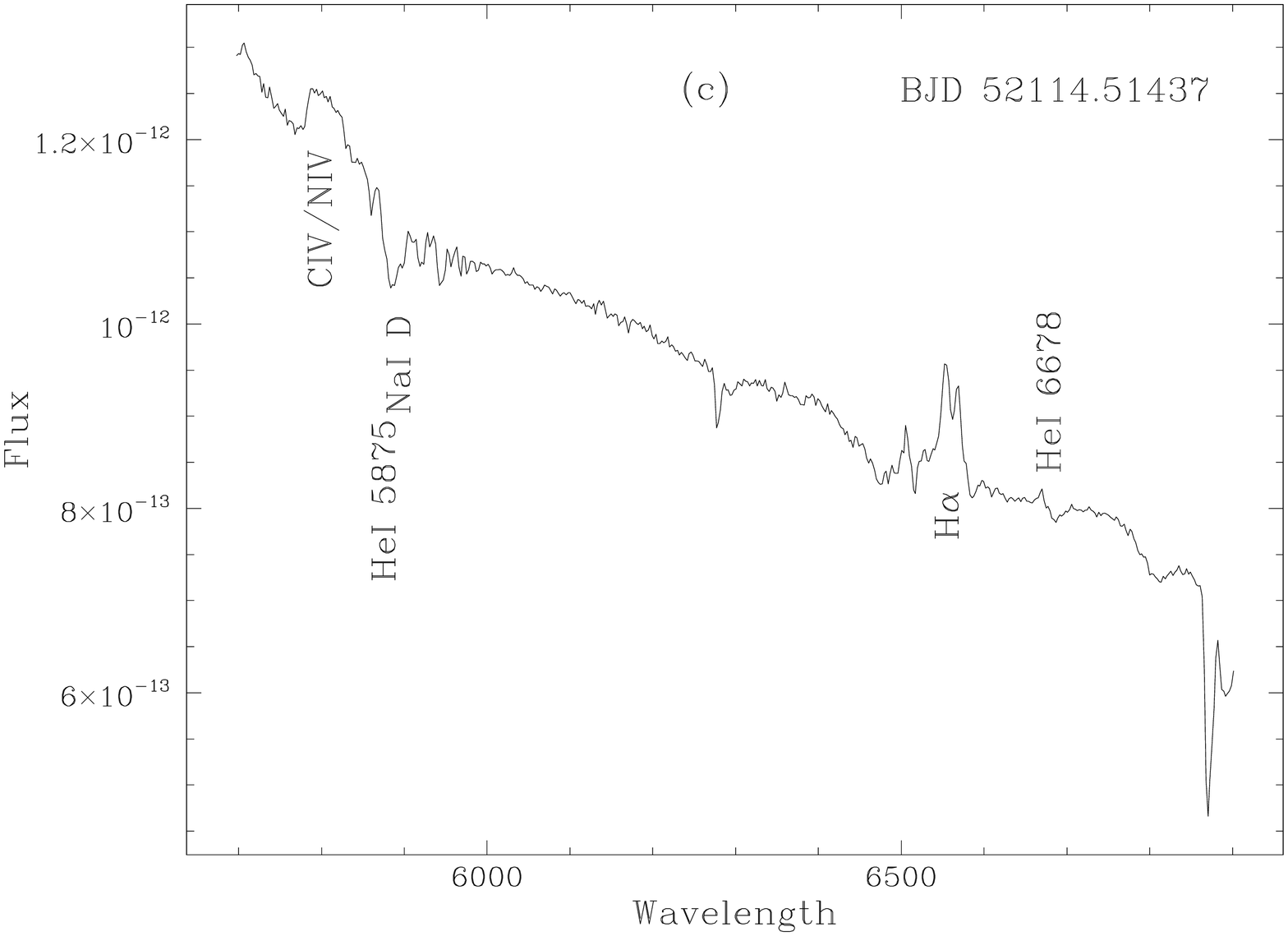}
  \end{center}
    \caption{Spectra (File ID: 10550, 10553, and 10558) around the
    maximum of the 2001 outburst of WZ Sge.  The abscissa is
    the wavelength in \AA\ and the ordinate is the flux in erg
    cm$^{-2}$ s$^{-1}$ \AA$^{-1}$. The spectra have a very blue
    continuum and show Balmer, He~\textsc{i}, and Na \textsc{i} D
    absorption lines, and He~\textsc{ii}, C~\textsc{iii}/N~\textsc{iii}
    Bowen blend, C~\textsc{iv} and N~\textsc{iv} emission lines.}
  \label{fig:firstspectra}
\end{figure}

\begin{figure}
  \begin{center}
    \FigureFile(84mm,115mm){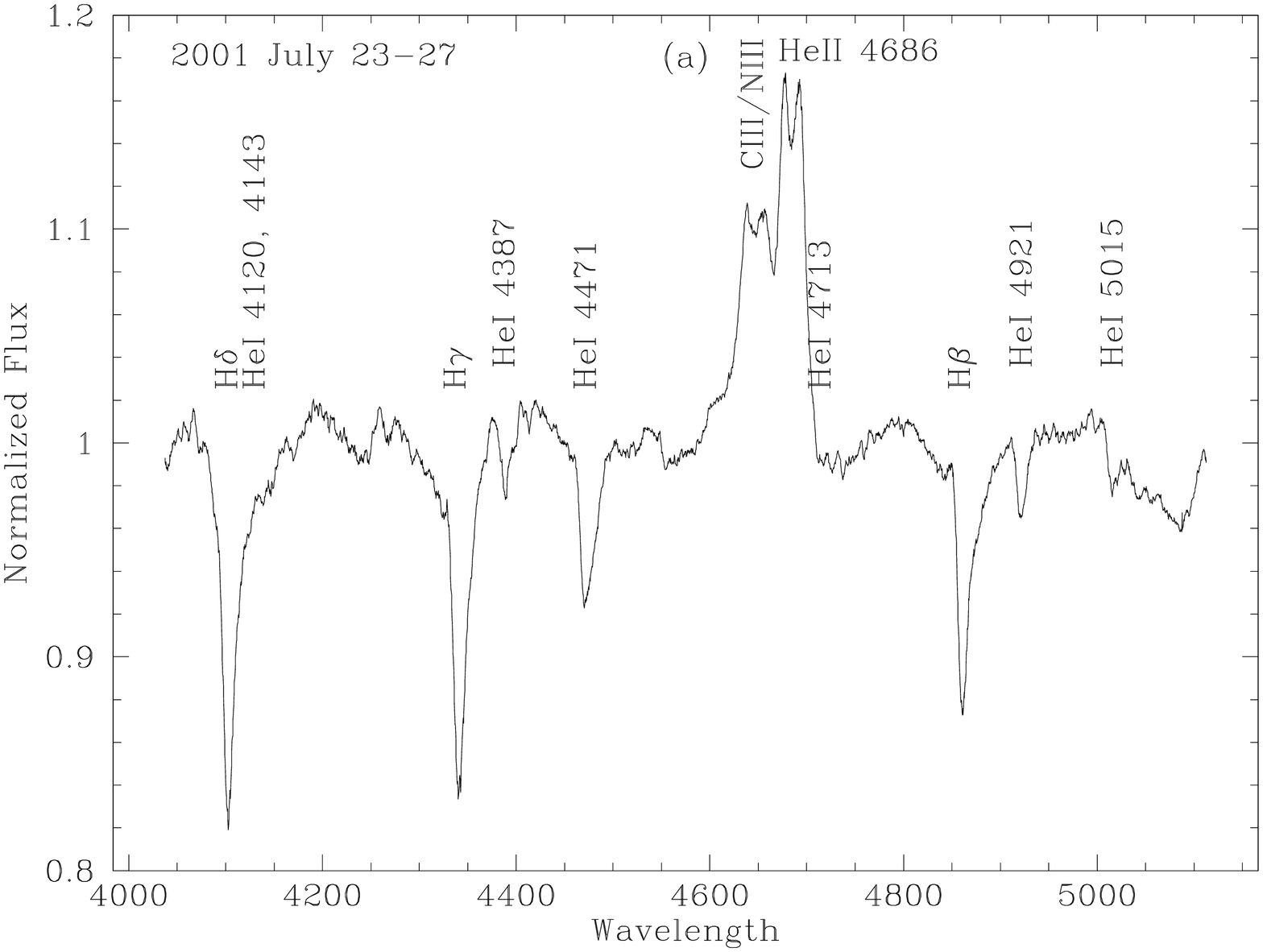}
    \FigureFile(84mm,115mm){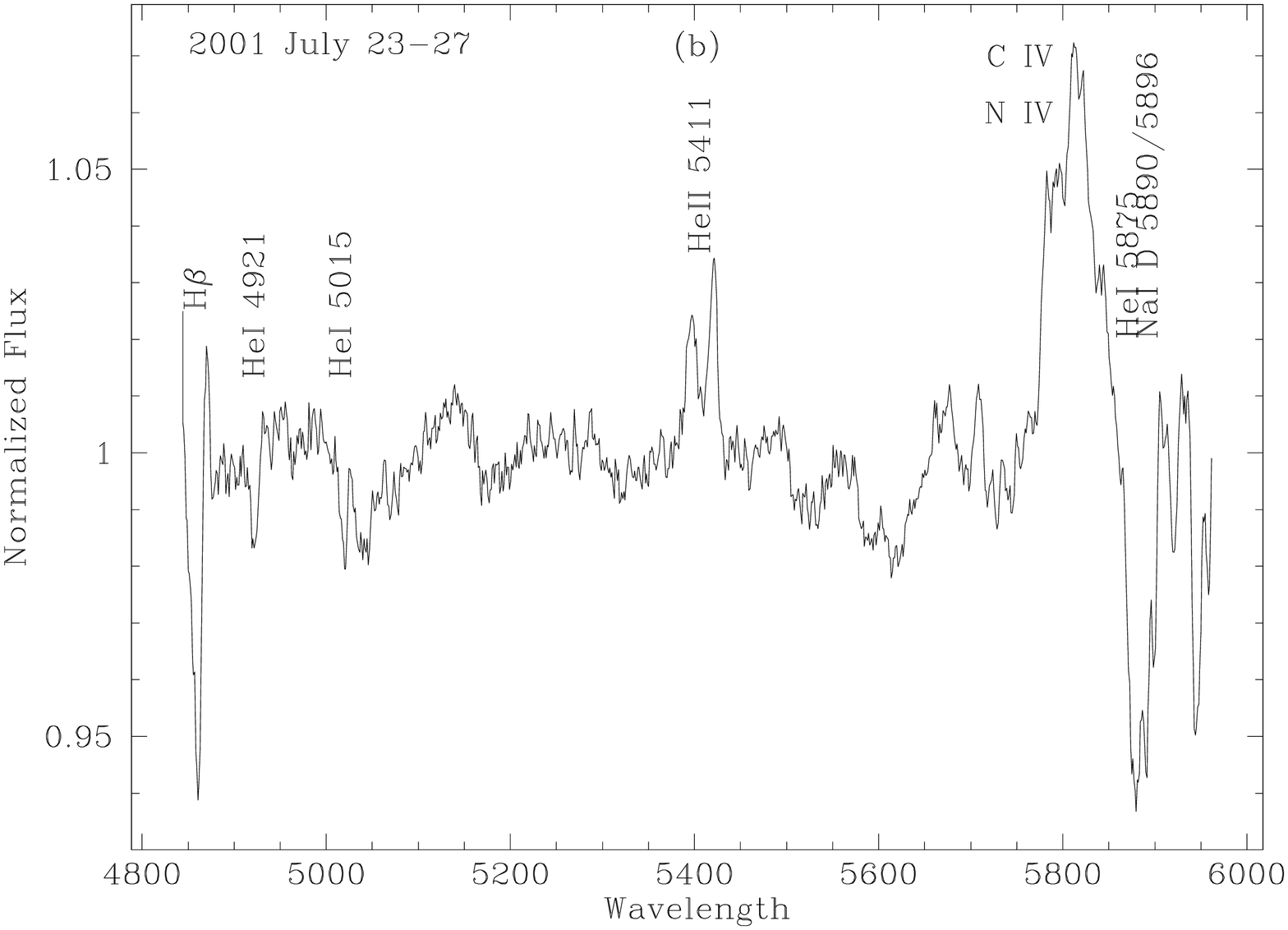}
    \FigureFile(84mm,115mm){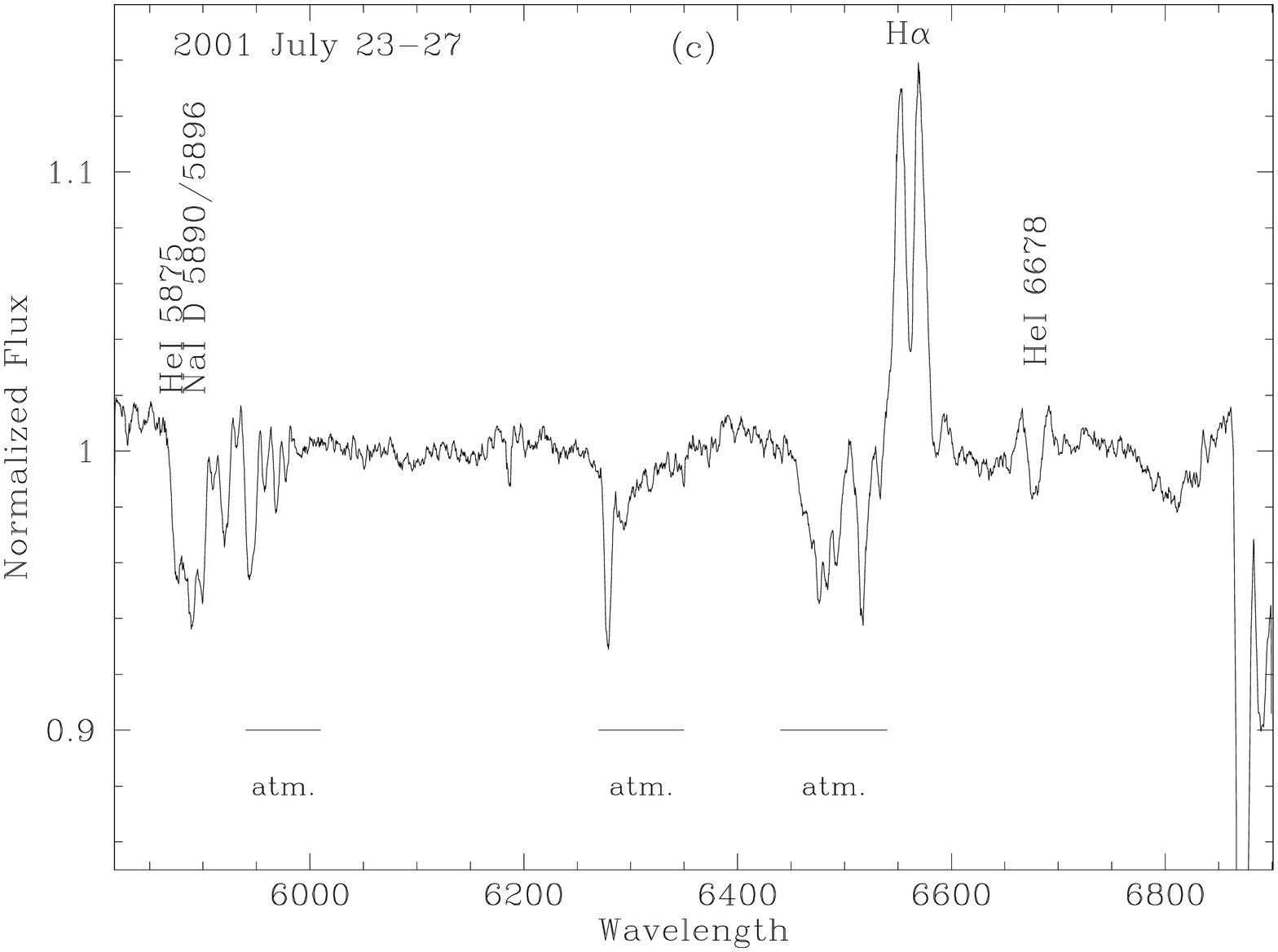}
  \end{center}
    \caption{Normalized spectra obtained by averaging the period I data
 of File ID: 10550--10603.  Balmer, other than H$\alpha$, and
 He~\textsc{i} lines are in absorption, and H$\alpha$ and He~\textsc{ii}
 emission lines have a doubly-peaked shape.  Other high-excitation
 lines, i.e. C~\textsc{iii}/N~\textsc{iii}, C~\textsc{iv}/N~\textsc{iv},
 are also present.  The broad absorption feature around 5600\AA\ in
 Figure b and the broad and narrow troughs on the bluer side of
 H$\alpha$ in figure c are due to the atmospheric absorption.
    }
  \label{fig:Isp}
\end{figure}

\subsection{Period I (JD 2452114-2452118, 1st-5th day)}

The spectra obtained on JD 2452114 just around the outburst
maximum are shown in figure \ref{fig:firstspectra}.  The continuum flux
in figure \ref{fig:firstspectra}b seems higher than those in Figure
\ref{fig:firstspectra}a and \ref{fig:firstspectra}c.  This discrepancy
should represent intrinsic modulations of the continuum, since the
amplitude of the early superhump were about 0.5 mag around those
observations \citep{ish04wzsge}.  The error of the flux calibration is
within 10 \%.

The continuum is very blue, and deep Balmer (except for H$\alpha$ in
emission) and He~\textsc{i} absorption lines are superposed on it, while
He~\textsc{ii} and the C~\textsc{iii}/N~\textsc{iii} Bowen blend are in
emission.  Note that the strong high-excitation emission lines
C~\textsc{iv} 5802/5813 and N~\textsc{iv} 5786/5796 are also present,
which were first caught in dwarf novae.

Figure \ref{fig:Isp} exhibits the spectra obtained by averaging all
the spectra obtained during the period I after normalization at the
continuum level.  H$\alpha$ and He~\textsc{ii} are in doubly-peaked
emission.  The peak separations in the spectra where the blue- and
red-peak wavelengths are easily measured by Gaussian fitting are
summarized in table \ref{tab:peak}.  Those of H$\alpha$ and
He~\textsc{ii} 4686 were rather small, $\sim$700 km~s$^{-1}$, in the
first night, then gradually increased.  The broadening rate was larger
in He~\textsc{ii} 4686 than in H$\alpha$.  In contrast, the separation
of He~\textsc{ii} 5411 was large since its emergence.  The
C~\textsc{iv} and N~\textsc{iv} emission lines were observed on JD
2452114 and 2452116.  These emission lines were, however, not detected,
or at least quite weak, on JD 2452128 and later.  Note that the
Na~\textsc{i} D 5890/5896 absorption lines were also clearly present
from our first run, while it suffered from significant contamination of
He~\textsc{i} 5875. These high-excitation lines and Na~\textsc{i}
features do not exist in optical quiescence spectra (see
e.g. \cite{gil86wzsge}).

\begin{table}
\caption{Peak separations (km s$^{-1}$) of emission lines averaged
each day.  The standard deviations are written in the parentheses.  Each
 observing period is partitioned by horizontal lines.}
\label{tab:peak}
\begin{center}
\begin{tabular}{crrrr}
\hline\hline
JD$^*$    & \multicolumn{1}{c}{He~\textsc{ii}} &
 \multicolumn{1}{c}{H$\beta$} & \multicolumn{1}{c}{He~\textsc{ii}} &
 \multicolumn{1}{c}{H$\alpha$} \\
 & \multicolumn{1}{c}{4686}  & & \multicolumn{1}{c}{5411} & \\ \hline
52114 & 660$^\dagger$ & & 1320$^\dagger$ & 740(40) \\
52116 & 1080(130) &  & 1390$^\dagger$ & 870(30) \\
52118 & 1050(150) &  &  & 840(100) \\ \hline
52119 & 1020$^\dagger$ &  &     \\
52121 & 1110(170) &  &  \\
52122 & 1210(140) &  &  & 1180(100) \\ \hline
52128 & 1260(230) & 1250(210) & & 1060(90) \\ \hline
52137 &  & 1370(90) & & \\
52138 &  & 1330(80) & & \\ \hline
52149 &  & 1370(40) & & 1300$^\dagger$ \\ \hline
52158 &  & &  \\
\hline
\multicolumn{5}{l}{$^*$JD$-$2400000.} \\
\multicolumn{5}{l}{$^\dagger$single spectrum.  The typical 1-$\sigma$
 error for one} \\
\multicolumn{5}{l}{measurement is $\sim$30 km s$^{-1}$.}
\end{tabular}
\end{center}
\end{table}

The daily averaged line fluxes and equivalent widths (EWs) of emission
lines (C~\textsc{iii}/N~\textsc{iii}, He~\textsc{ii} 4686,
C~\textsc{iv}/N~\textsc{iv} complex, and H$\alpha$) are listed in table
\ref{tab:flux}.  It is difficult to separate
C~\textsc{iii}/N~\textsc{iii} and He~\textsc{ii}, as seen in figure
\ref{fig:Isp}.  In this measurement,  we simply accumulated the flux in
the bluer region of the trough between the red peak of
C~\textsc{iii}/N~\textsc{iii} and the blue peak of He~\textsc{ii} as the
C~\textsc{iii}/N~\textsc{iii} line flux and in the redder region for
He~\textsc{ii}.  As for the C~\textsc{iv}/N~\textsc{iv} complex, we
measured the combined flux.  The line flux of the
C~\textsc{iii}/N~\textsc{iii} and He~\textsc{ii} lines became largest 4
days after the outburst maximum, while H$\alpha$ is almost constant.
The C~\textsc{iv}/N~\textsc{iv} complex was seen only first three days.

Table \ref{tab:ew} summarizes the daily averaged equivalent widths (EW)
of other Balmer and He~\textsc{i} absorption lines.  Some of them became
emissions later.

\subsection{Period II (JD 2452119-2452122, 6th-9th day)}

As in the period I, He~\textsc{ii} 4686, C~\textsc{iii}/N~\textsc{iii}
Bowen blend, and H$\alpha$ were doubly-peaked emission lines (figure
\ref{fig:IIsp}a).  The fluxes of C~\textsc{iii}/N~\textsc{iii},
He~\textsc{ii}, and H$\alpha$ were smaller during the period II than
those during the period I.  The peak separation of He~\textsc{ii} 4686
was about same as that at the end of the period I.  H$\alpha$ had peaks
separated significantly broader than in the period I, and the peak
separations of He~\textsc{ii} 4686 and H$\alpha$ were almost same.

The Balmer and He~\textsc{i} absorption lines started to exhibit
emission components in the wings (figure \ref{fig:IIsp}b).  Table
\ref{tab:ew} shows that the EWs decreased, but these absorption feature
became deeper between figures \ref{fig:Isp} and \ref{fig:IIsp}.  The
He~\textsc{ii} 4686 emission line suffered from significant
contamination by He~\textsc{i} 4713.

\begin{table*}
\caption{Line fluxes in a unit of 10$^{-12}$ erg cm$^{-2}$ s$^{-1}$
 \AA$^{-1}$ and equivalent widths in \AA (a negative EW for an emission
 line) of emission lines averaged each day.  The 1-$\sigma$ errors are
 written in parentheses.}\label{tab:flux}
\begin{center}
\begin{tabular}{crrrrr}
\hline\hline
JD$^*$& C \textsc{iii}       & He \textsc{ii}$^\dagger$ & He \textsc{ii} & C \textsc{iv}  & H$\alpha$ \\
      &  N \textsc{iii} & 4686 &  5411  & N \textsc{iv} \\ \hline
52114 & 7.2(0.9) & 6.6(0.8) & 1.6(0.2) & 4.5(0.8) & 2.6(0.3) \\
 (EW) & $-$2.8(0.1) & $-$2.7(0.1) & $-$0.8(0.1) & $-$3.6(0.4) & $-$3.1(0.1) \\
52116 & 5.9(1.0) & 6.7(1.0) & 1.1(0.2) & 2.7(0.3) & 2.4(0.3) \\
 (EW) & $-$3.8(0.3) & $-$4.0(0.2) & $-$0.8(0.1) & $-$3.3(0.1) & $-$3.2(0.1) \\
52118 & 7.7(1.5) & 8.4(1.1) & & & 2.4(1.0) \\
 (EW) & $-$5.9(0.6) & $-$6.5(0.2) & & & $-$4.1(1.3) \\ \hline
52119$^\ddagger$ &     &   &  &     \\
 (EW) & $-$3.1(0.2) & $-$4.5(0.2) & &  \\
52121 & 1.8(0.5) & 3.7(1.2) &  \\
 (EW) & $-$1.7(0.4) & $-$3.9(0.9) & \\
52122 & 2.8(0.6) & 3.7(0.6) & & & 1.1(0.3) \\
 (EW) & $-$3.3(0.4) & $-$4.7(0.4) & & & $-$3.9(0.7) \\ \hline
52128 & 0.8(0.2) & 2.6(0.5) & & & 2.4(0.8) \\
 (EW) & $-$1.2(0.2) & $-$4.7(0.5) & & & $-$8.2(2.0) \\ \hline
52137 & \multicolumn{2}{c}{1.0(0.2)$^\S$} &   \\
 (EW) & \multicolumn{2}{c}{$-$1.1(0.1)$^\S$} &   \\
52138 & \multicolumn{2}{c}{0.6(0.2)$^\S$} &   \\
 (EW) & \multicolumn{2}{c}{$-$0.8(0.1)$^\S$} &   \\ \hline
52149 &   --      &   --     & &  & 0.5(0.1) \\
 (EW) &   --      &   --     & &  & $-$15.5(0.1) \\ \hline
52158 &   --      &   --      &        \\
 (EW) &   --      &   --      &        \\
\hline
\multicolumn{6}{l}{$^*$JD$-$2400000.} \\
\multicolumn{6}{l}{$^\dagger$Contaminated by the He~\textsc{i} 4713
 absorption line.} \\
\multicolumn{6}{l}{$^\ddagger$Flux was not able to be measured due to
 clouds.} \\
\multicolumn{6}{l}{$^\S$Combined flux of
 C~\textsc{iii}/N~\textsc{iii} and He~\textsc{ii}.} \\
\end{tabular}
\end{center}
\end{table*}

\begin{table*}
\caption{Line fluxes in a unit of 10$^{-12}$ erg cm$^{-2}$ s$^{-1}$
 \AA$^{-1}$ and equivalent widths in \AA (a negative EW for an emission
 line) of absorption lines averaged each day.
}\label{tab:ew}
\begin{center}
\begin{tabular}{crrrrrrr}
\hline\hline
JD$^*$& He~\textsc{i}&  H$\delta$ &  H$\gamma$ & He~\textsc{i} &
 He~\textsc{i}&  H$\beta$ & He~\textsc{i} \\
      & 4026 & 4101 & 4340 & 4388 & 4471 & 4861 & 4922 \\ \hline
52114 & $-$3.3(0.3)& $-$12.4(2.0)& $-$11.1(1.1)& $-$0.6(0.2) &
 $-$4.8(0.3) & $-$7.4(0.2) & $-$1.0(0.2) \\
(EW)  & 1.0(0.1) &  4.0(0.1) &  4.0(0.1) &  0.2(0.1) & 1.8(0.1) &
 3.5(0.1) &  0.5(0.1) \\
52116 & $-$1.5(0.5)& $-$9.6(2.9) & $-$7.7(1.7) & $-$0.8(0.4) &
 $-$2.7(0.6)& $-$3.9(2.4) & $-$0.5(0.3) \\
(EW)  & 0.6(0.2) &  4.0(1.2) &  3.7(0.8) &  0.4(0.2) & 1.4(0.3) &
 2.5(1.5) &  0.3(0.2) \\
52118 & $-$1.2(0.4)& $-$5.8(2.5) & $-$4.4(1.4) & $-$0.7(0.2) & $-$2.4(0.8)&
 $-$2.2(1.3) & $-$0.6(0.3) \\
(EW)  & 0.6(0.2) &  3.0(1.3) &  2.6(0.8) &  0.4(0.1) & 1.5(0.5) &
 1.7(1.0) &  0.5(0.2) \\
52119$^\ddagger$  \\
(EW)  & 0.5(0.1) &  3.4(0.1) &  3.1(0.1) &  0.4(0.1) & 1.6(0.1) &  2.0(0.1) &  0.5(0.1) \\
52121 & $-$0.9(0.3)& $-$4.5(1.3) & $-$3.5(1.3) & $-$0.3(0.1) & $-$1.5(0.4)&
 $-$1.4(0.3) & $-$0.4(0.2) \\
(EW)  & 0.7(0.2) &  3.5(1.0) &  3.1(1.1) &  0.3(0.1) & 1.4(0.4) &  1.7(0.4) &  0.5(0.2) \\
52122 &     & $-$3.1(0.5) & $-$2.9(0.4) & $-$0.4(0.2) & $-$1.6(0.3)& $-$1.3(0.2) &
 $-$0.5(0.2) \\
(EW)  &     &  2.6(0.4) &  2.7(0.4) &  0.4(0.2) & 1.6(0.3) &  1.6(0.2) &
 0.6(0.2) \\ \hline
52128 &     &      &      & $-$1.0(0.3) & $-$1.1(0.3)& $-$0.6(0.6) &
 0.05(0.05) \\
(EW)  &     &      &      &  1.7(0.6) & 1.9(0.5) &  1.3(1.2) &
 $-$0.1(0.1) \\ \hline
52137 &$-$0.08(0.03)& 0.03(0.03) & $-$0.10(0.03)& 0.05(0.02) & $-$0.1(0.1) &
 0.29(0.02) & 0.13(0.02) \\
(EW)  & 0.3(0.1) & $-$0.1(0.1) &  0.4(0.1) & $-$0.2(0.1) & 0.5(0.4) &
 $-$1.5(0.1) & $-$0.7(0.1) \\
52138& $-$0.06(0.08) & 0.27(0.15)& 0.49(0.10) & 0.09(0.03) & $-$0.03(0.05)&
 0.74(0.14) & 0.11(0.07) \\
(EW) & 0.3(0.4) & $-$1.4(0.8) & $-$2.9(0.6) & $-$0.5(0.2) & 0.2(0.3) &
 $-$5.9(1.1) & $-$0.9(0.6) \\ \hline
42149 &     &      & 0.17(0.02)  &      &     & 0.25(0.03)  & \\
(EW)  &     &      & $-$3.4(0.4) &  --  & --  & $-$6.2(0.8) &
 $-$0.1(0.1) \\ \hline
52158 & 0.09(0.04) & $-$0.57(0.03)& $-$0.53(0.10)&     & $-$0.13(0.01)&
 $-$0.27(0.03) \\
(EW)  & $-$0.7(0.3) &  4.4(0.2) &  4.4(0.8) &  --  & 1.1(0.1) &  2.8(0.3) \\
\hline
\multicolumn{8}{l}{$^*$JD$-$2400000.} \\
\multicolumn{8}{l}{$^\dagger$Contaminated by Na~\textsc{i}~D} \\
\multicolumn{6}{l}{$^\ddagger$Flux was not able to be measured due to
 clouds.} \\
\end{tabular}

\begin{tabular}{crr}
\hline\hline
JD    & He~\textsc{i} &  He~\textsc{i} \\
      & 5875$^\dagger$ & 6678 \\ \hline
52114 & $-$1.0(0.1) & $-$0.2(0.1) \\
(EW)  &  0.9(0.1) & 0.3(0.1) \\ 
52116 & $-$0.3(0.1) & 0.2(0.1) \\
(EW)  &  0.3(0.1) & $-$0.3(0.1) \\
52118 & $-$0.8(0.3) & 0.1(0.1) \\
(EW)  &  1.0(0.4) & $-$0.2(0.2) \\ \hline
52119 & \\
(EW)  &      & \\
52121 & \\
(EW)  & \\
52122 & $-$0.7(0.2) & 0.03(0.12) \\
(EW)  &  1.7(0.5) & $-$0.1(0.4) \\ \hline
52128 & $-$0.5(0.2) & 0.3(0.1) \\
(EW)  &  1.6(0.7) & $-$1.2(0.5) \\ \hline
52137 & \\
(EW)  & \\
52138 & \\
(EW)  & \\ \hline
52149 & 0.02(0.00) \\
(EW)  & $-$0.7(0.1) \\ \hline
52158 & \\
(EW)  & \\ \hline
\end{tabular}

\end{center}
\end{table*}

We have performed period analyses on the line flux of emission lines
and the equivalent width of absorption lines obtained during this
period, in the range of 10--100 d$^{-1}$.  In these analyses, we used
the data obtained in BDJD 2452121.409--2452122.467 (July 30 and 31) for
H$\delta$, H$\gamma$, He \textsc{i} 4387, He \textsc{i} 4471, C
\textsc{iii}/N \textsc{iii}, He \textsc{ii} 4686, H$\beta$, and
He~\textsc{i} 4922, and the data obtained in HJD
2452122.506--2452122.607 (August 1) for He~\textsc{i} 5875, H$\alpha$,
and He~\textsc{i} 6678.  The power spectra are exhibited in figure
\ref{fig:ew}.  Most of the lines seem to have modulated with periods
close to $P_{\rm orb}$.  Confirmation is, however, needed, since the
coverages are not quite sufficient.  Note that \citet{how03wzsgeIR} also
reported EW modulations of H and He with respect to the orbital phase in
IR spectra obtained on 2001 July 27.

\subsection{Period III (JD 2452128, 15th day)}

The spectra were obtained with an echelle spectrograph in this period
(figure \ref{fig:IIIsp}).  While the emission line fluxes of He
\textsc{ii} 4686 and C \textsc{iii}/N \textsc{iii} were smaller than
those in the period II, that of H$\alpha$ become larger than in the
period II (table \ref{tab:flux}).  The intensities of the peaks of
H$\alpha$, especially the bluer peak, became much higher.  The peak
separations of He~\textsc{ii} 4686 and H$\alpha$ were 1260($\pm$230)
km~s$^{-1}$ and 1060($\pm$90) km s$^{-1}$, respectively (table
\ref{tab:peak}).

As listed in table \ref{tab:ew}, He~\textsc{i} absorption lines varied
differently from a line to a line.  He \textsc{i} 4388 became broader in
EW, but He \textsc{i} 4471 was almost constant from the period I.  He
\textsc{i} 4922 and 6678 was in emission.  The strong contamination from
Na~\textsc{i} makes it difficult to accurately measure the equivalent
width of He \textsc{i} 5875.  The H$\beta$ equivalent width became
narrower, and the H$\alpha$ emission line grew in EW.  The emission
components of He~\textsc{i} and Balmer lines got strong and the
absorption components became deeper since from period II.

\subsection{Period IV (JD 2452137-2452138, 24th-25th day)}\label{sec:IV}

This period corresponds to the very end of the main outburst.
On the second day of this period, JD 2452138, WZ Sge was in the rapid
fading phase.  The averaged, normalized spectrum is drawn in figure
\ref{fig:IVsp}.  The spectra obtained on JD 2452137 show H$\gamma$ and
H$\delta$ emission components possibly superposed on shallow absorption
features, while H$\beta$ is a pure, strong, doubly-peaked emission line
(table \ref{tab:ew}).  All the Balmer lines were in pure emission on JD
2452138.  He~\textsc{i} 4387 and 4471 were in absorption, but
He~\textsc{i} 4921 was in emission with a doubly-peaked shape.  The peak
separations of H$\beta$ were $\sim$1350 km s$^{-1}$.

The C~\textsc{iii}/N~\textsc{iii} and He~\textsc{ii} 4686 emission lines
had declined and was not able to be separated.  They were, however, still
present at the end of the main outburst. 

\begin{figure}
  \begin{center}
    \FigureFile(84mm,115mm){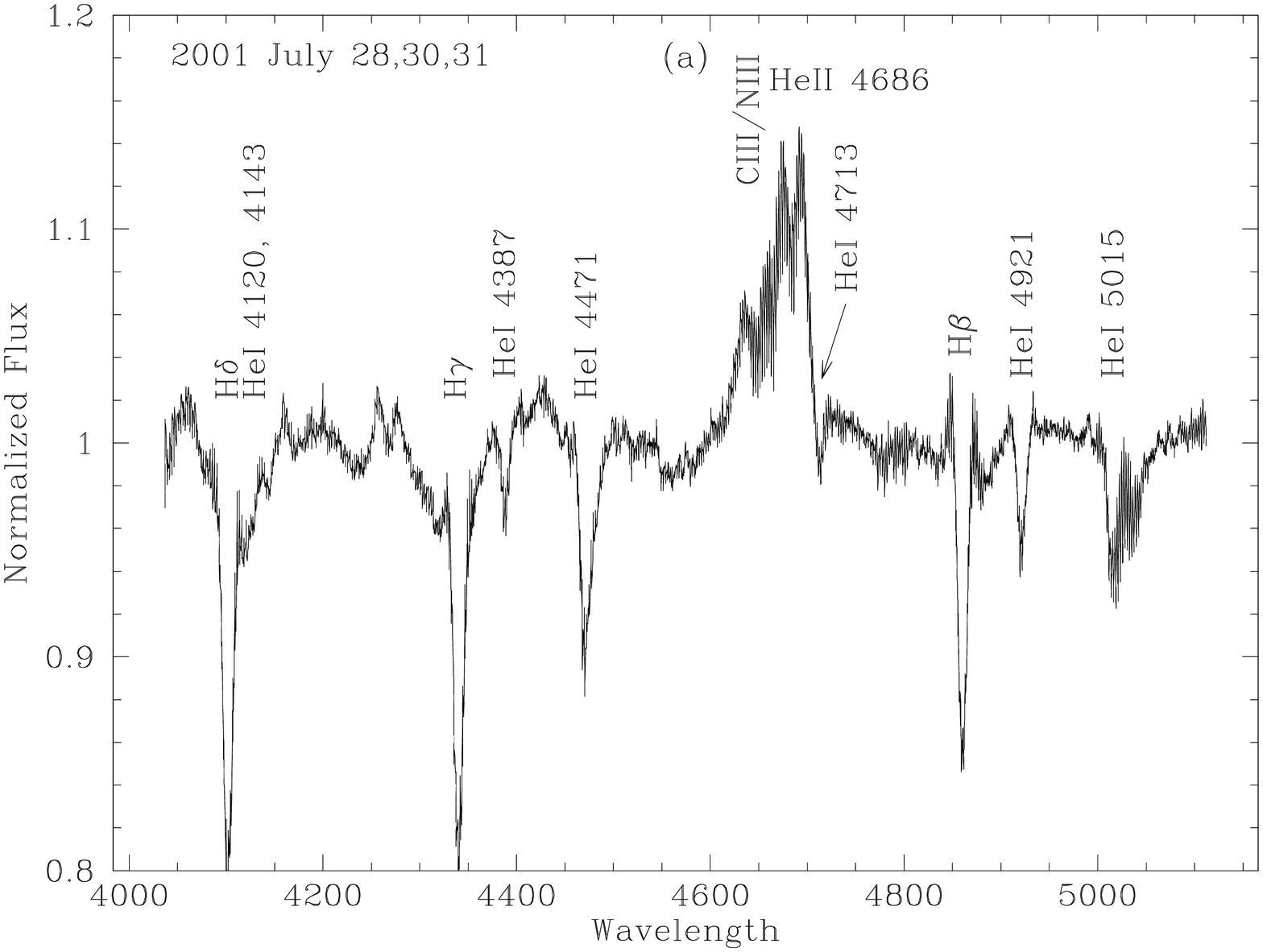} \\
    \FigureFile(84mm,115mm){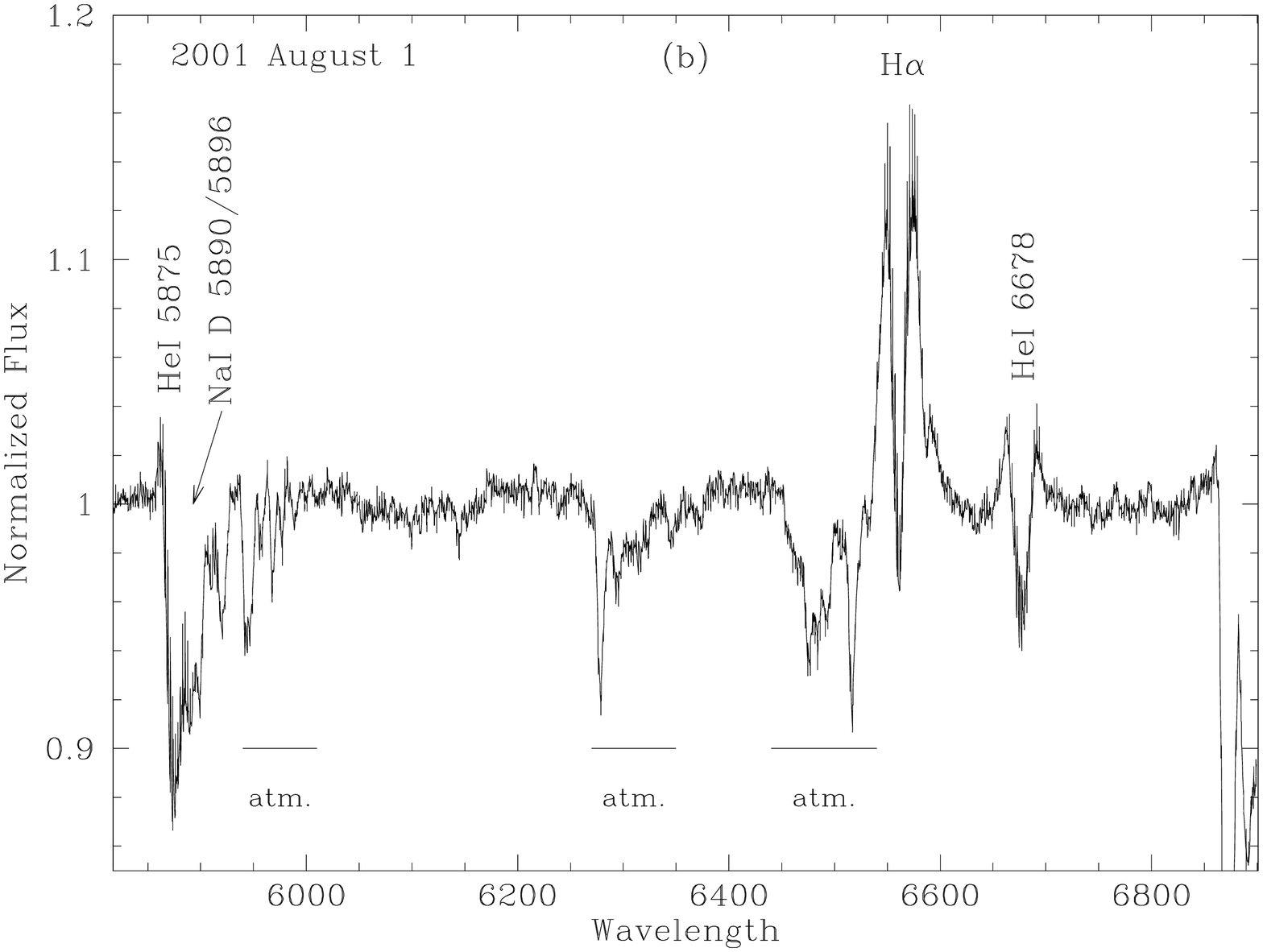}
  \end{center}
    \caption{Normalized spectra obtained by averaging the period II
    data of File ID: 10613--10679.  Balmer, other than H$\alpha$, and
    He~\textsc{i} lines are in absorption, while some of these lines
    exhibit emission components in the wing.  H$\alpha$ and
    He~\textsc{ii} emission lines have a doubly-peaked shape, as in
    figure \ref{fig:Isp}.
    }
  \label{fig:IIsp}
\end{figure}

\begin{figure}
  \begin{center}
    \FigureFile(84mm,115mm){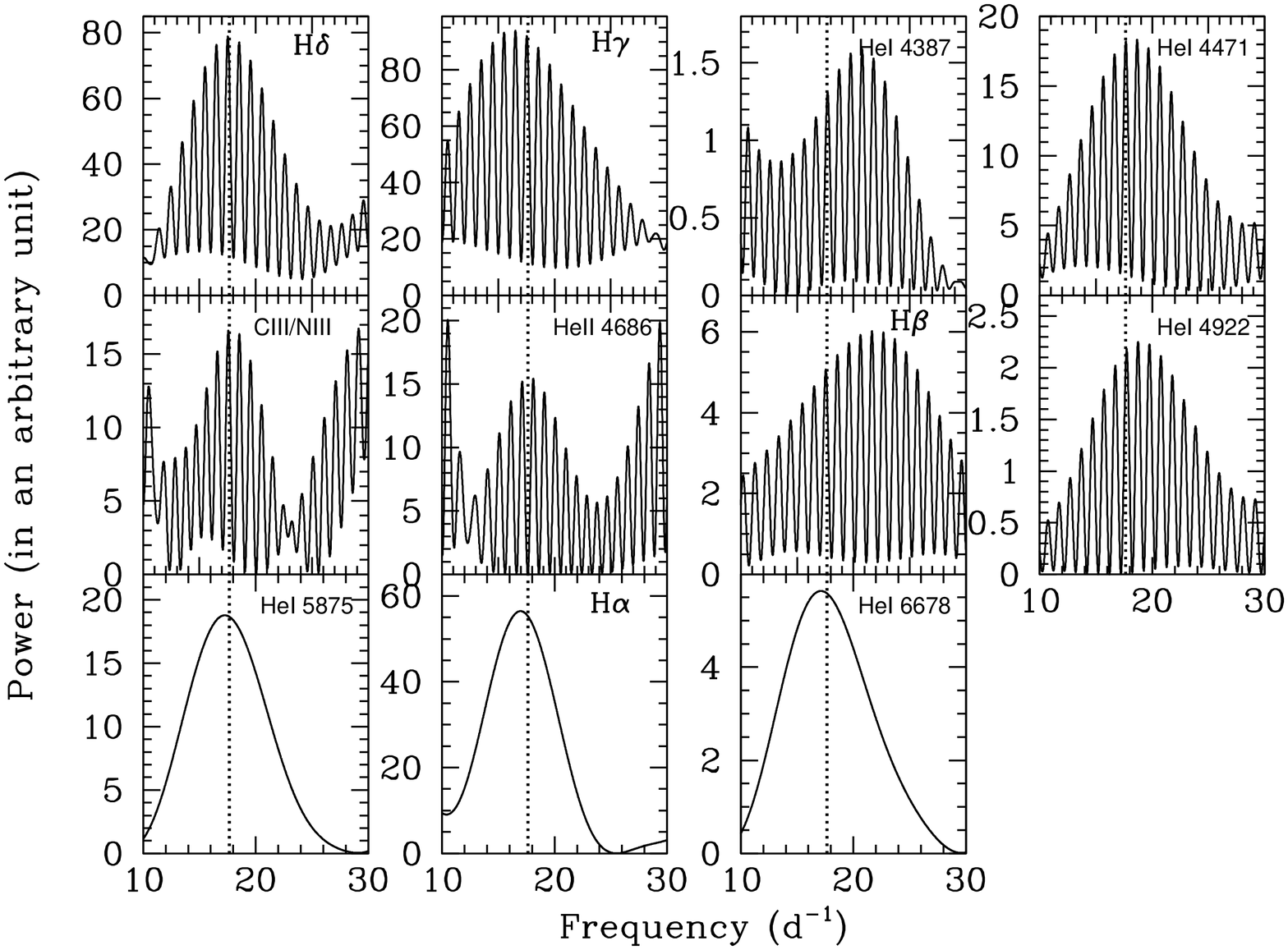}
  \end{center}
    \caption{Power spectra of the EW or line-flux variation in the
    period II.  The dotted lines represent the orbital frequency.  Most
    of lines seem to show modulations with periods close to
    $P_{\rm orb}$.
    }
  \label{fig:ew}
\end{figure}

\begin{figure}
  \begin{center}
    \FigureFile(84mm,115mm){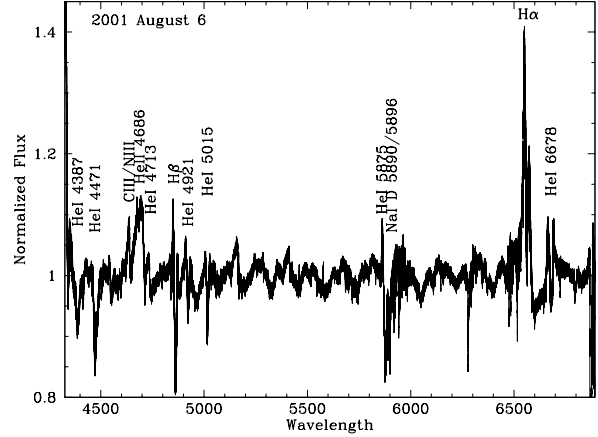}
  \end{center}
    \caption{Normalized spectra obtained by averaging the period III
    data of File ID: 37106--37122.  The small `waves' with a typical
    scale of $\sim$130 \AA\ are due to difficulty of complete
    sensitivity correction on echelle spectra.
    }
  \label{fig:IIIsp}
\end{figure}

\subsection{Period V (JD 2452149, 36th day)}

\begin{figure}
  \begin{center}
    \FigureFile(84mm,115mm){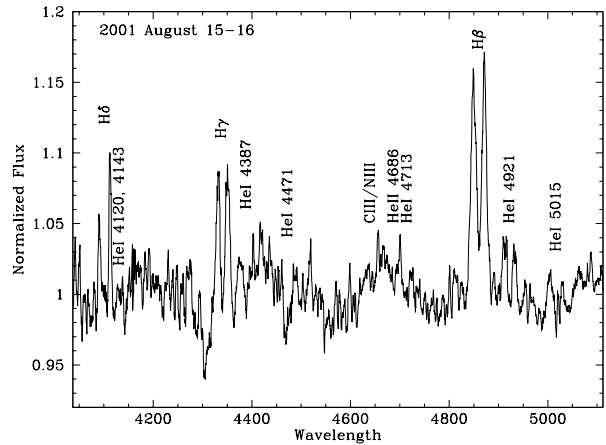}
  \end{center}
    \caption{Normalized spectrum obtained by averaging the period IV
    data of File ID: 10729--10746, at the end of the main outburst.
    Balmer lines became in strong emission.  He~\textsc{i} 4471 was a
    weak absorption, and He~\textsc{i} 4921 was a weak doubly-peaked
    emission line.  The He~\textsc{ii} 4686 and
    C~\textsc{iii}/N~\textsc{iii} emission lines had decreased, but were
    still clearly detectable.
    }
  \label{fig:IVsp}
\end{figure}

\begin{figure}
  \begin{center}
    \FigureFile(84mm,115mm){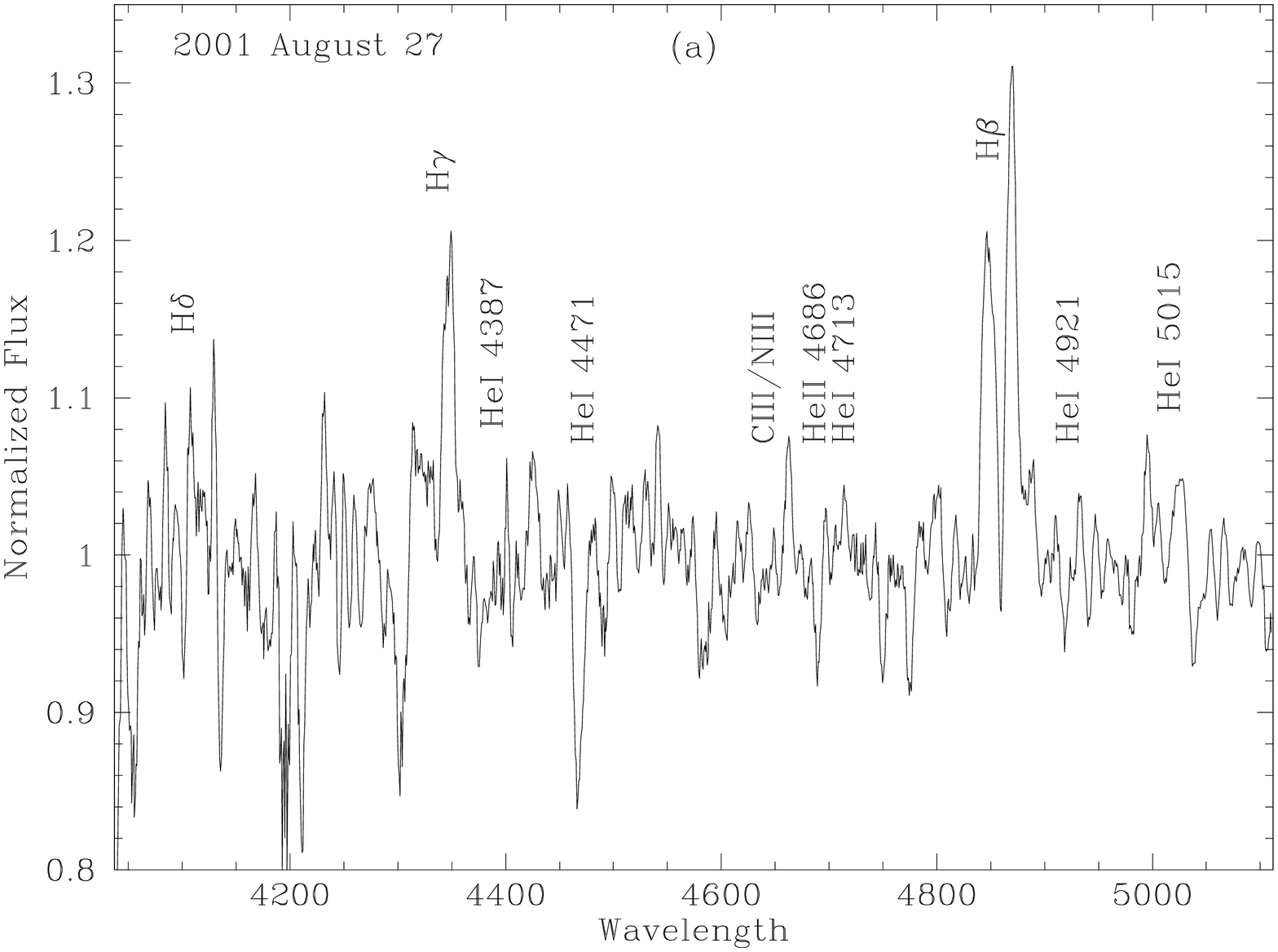}\\
    \FigureFile(84mm,115mm){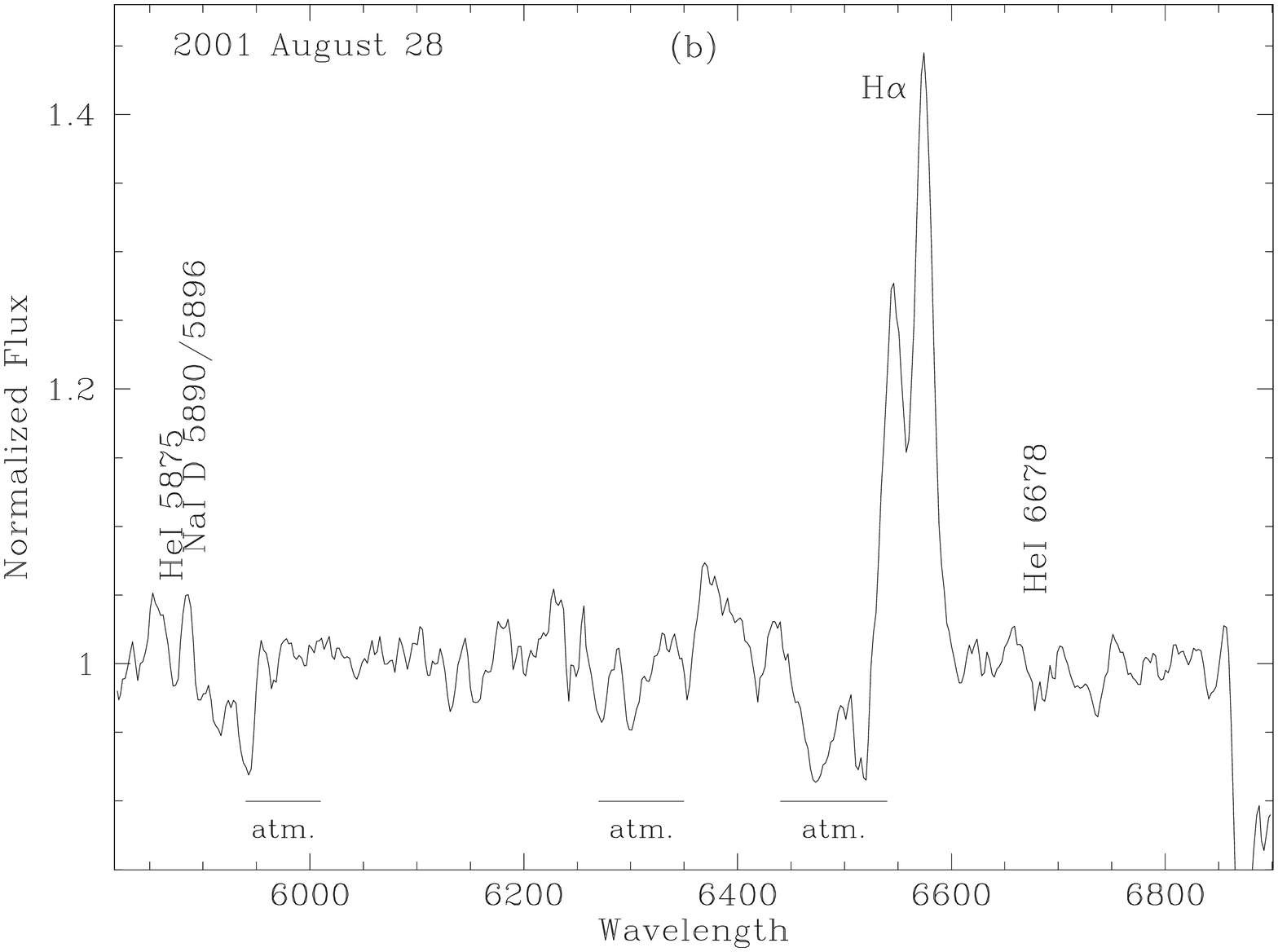}
  \end{center}
    \caption{Normalized spectrum obtained by averaging the period V
    data of File ID: 10993--10997, and normalized spectrum of File ID:
    11003, which were obtained during the dip between the third and
    the fourth rebrightenings.  Spectral feature in the upper panel
    was almost same as in figure \ref{fig:IVsp}.  He~\textsc{ii} 4686
    and C~\textsc{iii}/N~\textsc{iii} were not detectable.
    Na~\textsc{i} D doublet observed during the main outburst
    disappeared.
    }
  \label{fig:Vsp}
\end{figure}

As stated above, WZ Sge repeated 12 short rebrightenings, after the main
outburst.  The observation of this period were carried out around the
bottom of the dip between the third and fourth rebrightenings ($V \sim
12.4$).  Spectral feature in figure \ref{fig:Vsp}a is almost identical
with that in the period IV (figure \ref{fig:IVsp}).  The He~\textsc{ii}
4686 and C~\textsc{iii}/N~\textsc{iii} emission lines, however,
disappeared.  In figure \ref{fig:Vsp}b, He~\textsc{i} 5875 is a
doubly-peaked emission line and the Na~\textsc{i} D doublet which had
been in deep absorption during the main outburst was not detectable.
The peak separation of H$\alpha$ and H$\beta$ were $\sim$1300
km~s$^{-1}$ and 1370(40) km s$^{-1}$, respectively.  The red components
of the doubly-peaked shapes in Balmer emission lines have stronger peak
intensities than the blue components, while both components have almost
same intensity in He~\textsc{i} 5875.

\subsection{Period VI (JD 2452158, 45th day)}

\begin{figure}
  \begin{center}
    \FigureFile(84mm,115mm){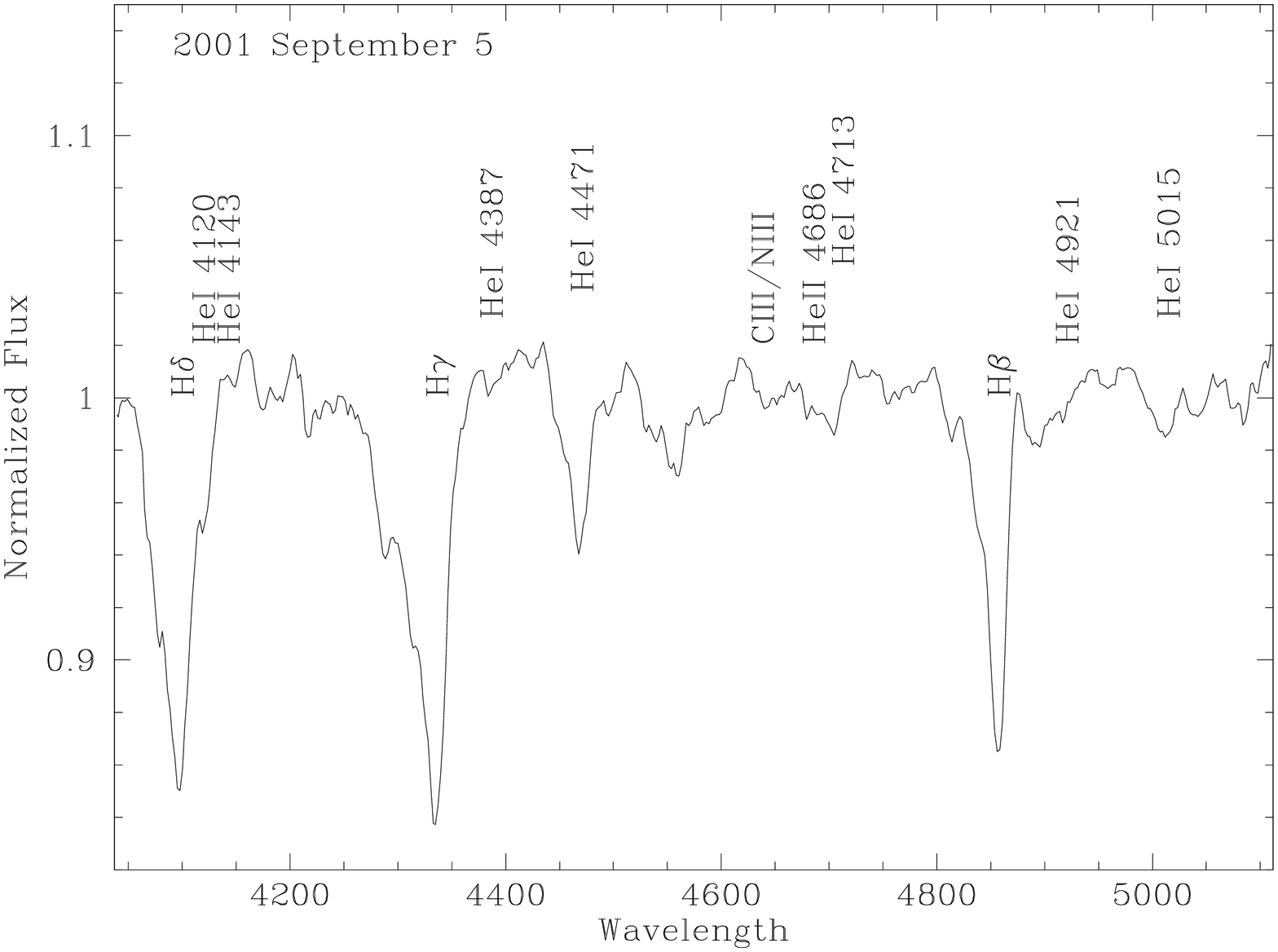}
  \end{center}
    \caption{Normalized spectrum of File ID: 11021.  This spectrum was
    obtained at the 9th peak (period VI).
    }
  \label{fig:VIsp}
\end{figure}

Our last observation was performed around the top of the 9th
rebrightening ($V \sim 11.7$).  All the Balmer lines were in strong
absorption (figure \ref{fig:VIsp}).  These absorption lines have
asymmetric shapes.  The red slopes may be affected by an emission
contribution.  Note that the full width at zero intensity (FWZI) is
broadest in H$\gamma$, and narrowest in H$\beta$ among the Balmer
lines. He~\textsc{i} 4471 was still in strong absorption, but the other
He~\textsc{i} lines were weak.  He~\textsc{ii} 4686 and
C~\textsc{iii}/N~\textsc{iii} were not detectable, as well as in the
period V.

\section{Spectral Evolution}

\subsection{The 2001 outburst}

The maximum time of this outburst can be accurately estimated to be
BDJD 2452114.7--2452114.8 with figure 1 in \citet{ish02wzsgeletter}.
The first 6 low-resolution spectra presented in \citet{bab02wzsgeletter}
were taken around HJD 2452114.26, about 12 hours before the maximum.
The average spectrum of these contains only absorption lines of
H$\alpha$, H$\beta$, and He~\textsc{i} 4922, 5876, and 6678.  We
performed the first run 5.5 hours after the observation by
\citet{bab02wzsgeletter}, but still before the outburst maximum.  As
shown in figure \ref{fig:firstspectra} and \ref{fig:Isp}, dramatic
changes occurred on the spectrum in the meantime: H$\alpha$ became an
emission line and high-excitation lines of C~\textsc{iii}/N~\textsc{iii}
blend, He~\textsc{ii} 4686 and 5411, and surprisingly, C~\textsc{iv}
5802/5813 and N~\textsc{iv} 5786/5796 emerged.  The
C~\textsc{iv}/N~\textsc{iv} complex has been caught for the first time
in the history of the dwarf nova study in the best knowledge of the
authors.  These C~\textsc{iv}/N~\textsc{iv} emission lines disappeared
by the 5th day of the outburst (figure \ref{fig:IIsp}).

The emission lines of He~\textsc{ii} 4686, 5411 and
C~\textsc{iii}/N~\textsc{iii} Bowen blend also emerged during 5.5 hours
before our 1st run.  In contrast to the C~\textsc{iv}/N~\textsc{iv}
complex, these emission lines increased their line flux in period I,
and persistently existed during the main outburst (table
\ref{tab:flux}).  \citet{kuu02wzsge} revealed existence of the spiral
structure in the emissivity distribution of He~\textsc{ii} 4686 and the
Bowen blend by the Doppler tomography technique
\citep{DopplerTomography} during the period I.  \citet{bab02wzsgeletter}
also reported an asymmetric spiral structure of He~\textsc{ii} 4686 and
no remarkable feature in the H$\alpha$ map.

The line flux of He~\textsc{ii} 4686 got larger for the period I, then
decreased till JD 2452121.  On the next day, He~\textsc{ii} 4686
rebrightened, and subsequently faded again.  We did not detect this
line during the rebrightening phase (period V and VI, figures
\ref{fig:Vsp} and \ref{fig:VIsp}).  The C~\textsc{iii}/N~\textsc{iii}
Bowen blend behaved in the almost same manner in the line flux.

The peak separation of He~\textsc{ii} 4686 was rather small, $\sim$660
km~s$^{-1}$ on JD 2452114, then kept almost constant around 1100
km~s$^{-1}$ during the periods I and II.  The small peak separation on
the first day, which was as small as that of H$\alpha$, may suggest that
the surface temperature was very high ($\geq$25,000 K) even at the edge
of the large accretion disk just before the outburst maximum
(cf. $\sim$1,400 km~s$^{-1}$ of the peak separation of Balmer lines in
quiescence, e.g. \cite{mas00wzsge}).  It is, however, noteworthy that
this small value was measured in a spectrum of File ID 10550 obtained at
an orbital phase of 0.255.  The trailed spectra of He~\textsc{ii} 4686
in \citet{bab02wzsgeletter} and \citet{kuu02wzsge} show that the peak
separation varied with the orbital phase and was smallest at the phase
around 0.25, although the peak separation might get broadened in the
period of $\sim$0.5 day between our first observation and that of
\citet{bab02wzsgeletter}.

The He~\textsc{ii} 5411 already had a large separation of over 1,300
km~s$^{-1}$ during the period I, which is as large as that of
He~\textsc{ii} 4686 on JD 2452128.  We can not see this line in the
spectra in the period III (figure \ref{fig:IIIsp}).

The Na \textsc{i} D absorption line was in deep absorption from the
period I to, at least, the period III (figures \ref{fig:Isp},
\ref{fig:IIsp}, \ref{fig:IIIsp}), although it was not seen during
the rebrightening phase (period V, figure \ref{fig:Vsp}).

At last, we describe the spectral evolution of Balmer lines.  The
spectra obtained by \citet{bab02wzsgeletter} around BDJD 2452114.26
demonstrated H$\alpha$ and H$\beta$ in absorption.  About 5.5 hours
later, however, H$\alpha$ was observed to be a strong, doubly-peaked
emission with a line flux of 2.6 $\times$ 10$^{-12}$ erg cm$^{-2}$
s$^{-1}$ \AA$^{-1}$, and an EW of $-$3.1 \AA (table \ref{tab:flux}).
These values are about 30 times larger than that in quiescence in the
line flux \citep{mas00wzsge} and about 30 times narrower than that in
quiescence in the EW \citep{gil86wzsge}.  The peak separation was
740(40) km s$^{-1}$, half of that in quiescence.  This fact suggests
that the radius of the region forming the H$\alpha$ emission line was
much larger just before the outburst maximum than that in quiescence.
This topic will be discussed later (section \ref{sec:Rmax}).  All the
higher Balmer lines were in absorption in the period I (figure
\ref{fig:Isp}).

With the outburst going, the H$\alpha$ emission line steadily became
strong in the EW, but stayed almost constant in the flux during the main
outburst.  The peak separation gradually grew.  Nevertheless, it was
about 1,100 km~s$^{-1}$ in the period III, still smaller than that in
quiescence.  In the period V, i.e. in a minimum in the rebrightening
phase, the flux of H$\alpha$ was smaller than that in the main outburst,
but still several times larger than that in quiescence.  The peak
separation was significantly larger than that in the main outburst, but
a little smaller than that in quiescence.

The other Balmer lines decreased their EWs with time, and then finally
became emission lines in the period III (H$\beta$) and in the period IV
(H$\gamma$ and H$\delta$), around the end of the main outburst.  The
H$\beta$ and higher Balmer lines were in absorption again in the period
VI.  The FWZIs of H$\gamma$ and H$\delta$ was much larger than those in
the period I, making the EWs in the period VI broader than those in the
period I.

\subsection{Comparison with the spectra in other outbursts in WZ Sge}

The spectra in the 15th, 17th, 19th, and 23rd day of the 1946
outburst observed by G. Herbig were reported by \citet{mcl53novaqui}.
These spectra had doubly-peaked emission lines of H$\beta$, H$\gamma$,
and H$\delta$ with peak separations of $\sim$15 \AA\ and FWZI of
$\sim$50 \AA, in contrast to that  H$\beta$ was in strong absorption
with weak doubly-peaked emission component on the 15th day in the
present outburst (period III, figure \ref{fig:IIIsp}).  The peak
separations of H$\beta$-H$\delta$ were $\sim$500--800 km s$^{-1}$ on the
15th day, much narrower than in the period III (section 4.3).   Wide,
weak absorptions of He~\textsc{i} 4026 and 4471 were also present.  They
noted detection of Fe \textsc{ii} 4233 and absorption features $-$2,400
km s$^{-1}$ apart from H$\gamma$ and H$\delta$ and $-$3,700 km~s$^{-1}$
apart from H$\beta$, which were not in our spectra.  On the other hand,
\citet{mcl53novaqui} mentioned that no other emission/absorption lines
were found, though our spectra on the 15th day of the 2001 outburst
(figure \ref{fig:IIIsp} and table \ref{tab:ew}) clearly show strong
He~\textsc{ii} emission line.

The 1978 outburst was noticed at 1978 December 1.1 (UT) and reached its
maximum on the same day \citep{pat78wzsgeiauc3311}.  The first
spectroscopic observation was started at December 1.7 (UT) by
\citet{bro80wzsgespec}.  H$\alpha$ had a weak, singly-peaked profile in
the spectra at December 1.7 (UT), but was observed to be in a
doubly-peaked shape with peak separations of $\sim$500 km s$^{-1}$ in
December 2.7--7.7.  This constancy of the peak separation of H$\alpha$
was independently revealed by \citet{gil80wzsgeSH}.  The velocities of
the peaks of H$\alpha$ were also constant, within $\pm$15 km~s$^{-1}$,
on December 7 \citep{gil80wzsgeSH}.

\citet{cra79wzsgespec} also obtained optical time-resolved spectra on
1978 December 6 and 7.  In their spectra, H$\beta$ was already in
emission with double peaks (peak separation $\sim$600 km~s$^{-1}$).
The semi-amplitudes of the radial velocities of H$\beta$ were small,
19($\pm$7) km~s$^{-1}$, 13($\pm$18) km~s$^{-1}$, and 43($\pm$9)
km~s$^{-1}$ for the central absorption, the blue peak, and the red peak,
respectively.  The average semi-amplitude of those radial velocities of
H$\gamma$--$\delta$ was larger, 162($\pm$37) km~s$^{-1}$.
C~\textsc{iii}/N~\textsc{iii} emission was already weak at that time.

\citet{pat78wzsgeiauc3311} reported on spectral lines in their
time-resolved spectra.  Although the date when the spectra were taken is
not written in the IAU Circular, the line feature of the Balmer series,
He~\textsc{i}, He~\textsc{ii}, and the Bowen blend is nearly identical
with ours in the period I and II.  The spectra in the early phase
obtained by \citet{ort80wzsge} also showed the same feature.

Nevertheless, the variation of the lines in the later phase in the main
outburst \citep{gil80wzsgeSH, ort80wzsge, wal80wzsgespec} was different
from that in our spectra: He~\textsc{ii} 4686 was already weak on the
10th day, and H$\alpha$ became weaker in contrast to that it became
stronger in our data (table \ref{tab:flux}).

In the dip and the rebrightening phase, the spectra in
\citet{ort80wzsge} was again the same as ours.

As seen in this section, the spectral feature and its evolution seem
slightly different among those in the 1946, 1978, and 2001 outbursts,
even in the same system WZ Sge.

\subsection{Comparison with the spectra in other SU UMa stars in
  superoutburst}

The spectroscopic observations of an eclipsing dwarf nova, Z Cha in
superoutburst were done by \citet{vog82zcha} and \citet{hon88zcha}.
Doubly-peaked emission lines of the Balmer series superposed on shallow,
broad absorption components, He~\textsc{i} absorptions, and emission
lines of He~\textsc{ii} 4686 and C~\textsc{iii}/N~\textsc{iii} were seen
in their spectra.  \citet{wu01iyuma} spectroscopically observed another
eclipsing star, IY UMa during the 2001 superoutburst.  They also
observed emission lines of the Balmer series, He \textsc{i}, and the
He~\textsc{ii}-C~\textsc{iii}/N~\textsc{iii} complex.  In addition,
Na~\textsc{i} D absorption was detected.

In contrast to the high-inclination systems including WZ Sge in this
paper, VY Aqr \citep{aug94vyaqr} and SX LMi \citep{wag98sxlmi} in
superoutburst did not have the emission line of He~\textsc{ii} 4686.
Another WZ Sge star, EG Cnc did not show the He~\textsc{ii} 4686
emission line \citep{pat98egcnc}.  Their spectra of EG Cnc were,
however, obtained during the rebrightening (echo-outburst) phase, and
this emission line was not firmly detected also in our spectra in the
rebrightening phase (phases V and VI, figures \ref{fig:Vsp} and
\ref{fig:VIsp}).  Na~\textsc{i} D absorption was seen in the EG Cnc
spectra, while we can not see it in the rebrightening phase (figure
\ref{fig:Vsp}).

The spectra of another WZ Sge star, V592 Her obtained on the 3rd day of
the outburst by \citet{men02v592her} showed absorption lines with an
emission core of the Balmer series including H$\alpha$ and
He~\textsc{i}, and a weak doubly-peaked emission line of He~\textsc{ii}
4686.  Na~\textsc{i} absorption was not present in the spectra.

In conclusion, visibility of the He~\textsc{ii} 4686 emission line seems
to depend on the inclination, and it is not only in WZ Sge-type dwarf
novae that the Na~\textsc{i} D line is seen.

\section{Discussion}

\subsection{Outburst properties}

The outbursts in WZ Sge had been observed three times in 1913, 1946, and
1978 with a recurrence cycle of 32.5 years.  Breaking this punctuality,
however, WZ Sge has underwent a new outburst {\it only} 23 years after
the last 1978 outburst.

Besides of the stable recurrence cycle of the first three outbursts,
the long term outburst light curve varied from one to the other (see
figure 1 in \cite{kuu02wzsge}; see also figure 1 in \cite{ort80wzsge}).
In the 1913 outburst, the main outburst seems to have lasted about
38 days, or maybe had lasted around 28 days and the point on the 37th
day from the onset was in the rebrightening phase (see
\cite{bro79wzsge}; \cite{ort80wzsge}).  The 1946 outburst consists
of only a main outburst of 29 days followed by a long fading tail
\citep{may46wzsge, esk63wzsge}.  During the 1978 outburst, the best
observed one among these three, WZ Sge had kept the plateau phase of the
main outburst for 32 days.  After a dip of $\sim$3 days following the
plateau phase, WZ Sge caused a rebrightening which lasted at least 20
days.   It is not clear whether this rebrightening was the `second
superoutburst' as in the AL Com 1995 outburst \citep{nog97alcom}, or
a complex of many repetitive outbursts like during the current
outburst.  Comparing these previous outbursts, the 2002 outburst has
characteristics of the relatively short main outburst of $\sim$25 days
and the following rebrightening phase consisting of 12 short outbursts
for 24 days.

The total energy emitted during the 1978 outburst seems a little larger
than that during the 2001 outburst, at least, if we compare them in
optical.  The ratio of them is, however, obviously smaller than the
ratio of the quiescence durations before each outburst, i.e. 33
years to 23 years.  Regarding the 1946 outburst and the 1978 one,
the outburst light curves indicate that the radiated energy during the
1946 outburst was much smaller than that during the 1978
outburst, while the quiescence durations before each outburst were
nearly the same, $\sim$33 years.  These imply that the condition to
cause an outburst in WZ Sge does not depend only on the mass stored in
the accretion disk, and the mass transfer changes even in WZ Sge (an
increasing trend at least in these several tens of years).

Variations of the outburst patters have been minutely demonstrated in
many dwarf novae, especially in recent years, such as in DI UMa
\citep{fri99diuma}, SU UMa \citep{ros00suuma, kat02suuma}, V1113 Cyg
\citep{kat01v1113cyg}, V503 Cyg \citep{kat02v503cyg}, DM Lyr
\citep{nog03dmlyr} and MN Dra \citep{nog03var73dra}.  The currently most
plausible model for these variations is rooted in the the solar-type
cycle in the secondary star \citep{war88CVcycle, bia90CVcycle,
ak01CVcycle}.  If this scenario is true, the secondary star in WZ Sge is
suggested to have magnetic activities, although the secondary is
considered to be (close to) a degenerate star as described above.

This would be observed as changes of the brightness in quiescence.
Although analyses of the quiescent eclipse times and the long-term light
curve of WZ Sge by \citet{ski04magneticactivity} have yielded no
evidence for cyclical modulations, it will be necessary to pile up the
quiescence observations for more decades in order to judge the magnetic
activity of the secondary star.  The continuous observation reports by
amateur observers will be of much help for this study.

\begin{table*}
\caption{Radial velocities of lines in a unit of km s$^{-1}$ on 2001
 July 30 and 31, and August 1 and 6.  The marks, $^*$ and $^{\dagger}$
 represent the blue component and the red component of the double peaks,
 respectively.  The typical 1-$\sigma$ error is $\sim$30 km s$^{-1}$.}
\begin{center}
\begin{tabular}{crrrrrrrrrr}
\hline\hline
July 30 & H$\epsilon$ & He~\textsc{ii} & H$\delta$ & H$\gamma$ & He~\textsc{i} & He~\textsc{ii}$^*$ & He~\textsc{ii}$^{\dagger}$ & H$\beta$ & He~\textsc{i} &
 He~\textsc{i} \\
Frame ID &     & 4026  &     &     & 4471 & 4686 & 4686 &   & 4921 & 5015
 \\ \hline
10616 & $-$6.9 & $-$44.0 & $-$43.9 & $-$58.8 & $-$88.2 & $-$855.3 & 582.9 & $-$52.0 &
 $-$58.1 & 126.2 \\
10618 & 161.3 & 257.6 & 47.7 & $-$15.6 & 63.9 & $-$469.8 & 614.1 & $-$3.7 &
 97.9 & 153.8 \\
10620 & 154.7 & 130.2 & 6.8 & $-$10.9 & 37.5 & $-$578.4 & 590.0 & $-$49.8 &
 50.5 & 68.7 \\
10622 & 75.3 & 119.8 & $-$49.6 & $-$88.8 & $-$120.8 & $-$467.5 & 482.2 & $-$90.8 &
 $-$59.8 & 150.1 \\
10624 & $-$19.5 & 47.7 & $-$94.8 & $-$118.5 & $-$96.2 & $-$865.8 & 312.5 & $-$132.1
 & $-$107.0 & 34.2 \\
10625 & $-$21.4 & 28.1 & $-$112.7 & $-$142.1 & $-$99.6 & $-$546.8 & 464.0 & $-$154.5
 & $-$47.9 & 28.3 \\
10626 & $-$12.5 & $-$24.2 & $-$78.8 & $-$159.1 & $-$119.6 & $-$568.1 & 453.2 &
 $-$137.3 & $-$70.6 & $-$102.7 \\
10628 & 47.8 & $-$19.8 & $-$90.5 & $-$86.3 & $-$158.2 & $-$859.1 & 474.9 & $-$62.2 &
 $-$185.9 & $-$87.1 \\
10629 & 125.4 & 209.2 & 49.0 & $-$53.6 & $-$9.7 & $-$570.5 & 486.9 & $-$44.0 &
 $-$18.2 & 105.9 \\
10630 & 276.7 &       & 185.9 & 145.6 & 185.0 & $-$344.5 & 577.2 & 106.1
 & 97.7 & 351.6 \\
10631 & 455.7 & 404.1 & 280.8 & 259.0 & 292.6 & $-$293.8 &       & 234.7 &
 360.8 & 415.5 \\
\hline
\\
\end{tabular}

\begin{tabular}{crrrrrrrr}
\hline\hline
July 31  & H$\delta$ & H$\gamma$ & He~\textsc{i} & He~\textsc{ii}$^*$ & He~\textsc{ii}$^{\dagger}$ & H$\beta$ & He~\textsc{i} &
 He~\textsc{i} \\
Frame ID &     &     & 4471 & 4686 & 4686 &   & 4921 & 5015 \\ \hline
10637 &    $-$6.4 & $-$105.5 &  $-$76.0 & $-$852.3 &  513.4 & $-$110.8 &
 $-$49.6 &  $-$91.4 \\			    
10639 &   $-$86.4 & $-$115.7 & $-$165.6 & $-$841.6 &  250.9 & $-$115.0 &
 $-$130.2 & $-$172.0 \\			    
10640 &   $-$11.5 &  $-$64.6 &  $-$14.0 & $-$753.4 &  469.5 &  $-$73.4 &
 $-$31.0 &   $-$7.0 \\			    
10642 &   179.2 &   79.3 &   99.8 & $-$432.2 &        &   73.6 &
 135.3 &   90.9 \\			    
10644 &   257.1 &  177.5 &  282.2 & $-$807.4 &  449.4 &  171.9 &
 174.4 &  270.0 \\			    
10645 &    83.7 &   40.3 &   94.3 & $-$752.6 &  673.0 &   20.8 &
 $-$41.9 &   67.1 \\			    
10646 &    79.5 &  $-$20.2 &  $-$32.7 & $-$543.7 &  501.7 &   $-$9.2 &
 $-$4.0 &  105.9 \\			    
10648 &   $-$57.5 & $-$111.0 & $-$122.0 & $-$678.1 &  358.0 & $-$137.1 &
 $-$77.6 &  $-$43.3 \\			    
10649 &  $-$128.0 & $-$175.8 & $-$176.9 & $-$829.3 &  376.0 & $-$168.7 &
 $-$127.7 & $-$151.8 \\			    
10650 &   $-$82.4 & $-$108.1 & $-$178.1 & $-$819.0 &  438.3 & $-$119.0 &
 $-$139.4 & $-$131.7 \\
\hline
\\
\end{tabular}

\begin{tabular}{crrrr}
\hline\hline
August 1 & He~\textsc{i} & H$\alpha^*$ & H$\alpha^{\dagger}$ & He~\textsc{i} \\
Frame ID & 5875 &         &                 & 6678 \\ \hline
10654 & 175.8   & $-$521.2 & 652.8 &  187.3 \\
10656 & 116.6   & $-$670.3 & 520.3 &  108.8 \\ 
10657 & 129.9   & $-$633.5 & 599.9 &   70.1 \\
10658 & $-$20.9 & $-$616.8 & 549.4 &  $-$26.7 \\
10659 & $-$72.9 & $-$618.8 & 369.6 &  $-$83.8 \\
10662 & $-$67.2 & $-$730.5 & 493.9 & $-$129.7 \\
10663 &  14.1   & $-$826.7 & 565.5 & $-$117.7 \\
10664 &  $-$5.7 & $-$700.7 & 494.7 &   $-$6.5 \\
10665 &  20.8   & $-$590.1 & 500.6 &  $-$45.7 \\
10667 & 136.3   & $-$508.5 & 539.0 &   54.0 \\
10668 & 251.4   & $-$496.0 & 674.4 &  190.7 \\
10669 & 142.3   & $-$510.7 & 663.2 &  131.7 \\
10670 & $-$39.5 & $-$636.6 & 670.7 &   45.5 \\
10672 &  28.8   & $-$656.1 & 590.3 &  $-$15.4 \\
10673 & $-$33.5 & $-$627.4 & 413.5 &  $-$37.9 \\
10674 & $-$49.4 & $-$637.4 & 479.0 &  $-$68.5 \\
10675 & $-$69.5 & $-$598.7 & 556.0 & $-$131.5 \\
10676 & $-$85.4 & $-$714.8 & 574.2 & $-$118.1 \\
10678 &  34.5   & $-$695.1 & 555.3 &   69.7 \\
10679 &  27.1   & $-$607.9 & 517.7 &   52.8 \\ \hline
\\
\end{tabular}
\end{center}
\end{table*}

\addtocounter{table}{-1}
\begin{table*}
\caption{(continued)}\label{tab:rv}
\begin{center}
\begin{tabular}{crrrrrrrrrr}
\hline\hline
August 6 & He~\textsc{i} & He~\textsc{i} & He~\textsc{ii}$^*$ & He~\textsc{ii}$^{\dagger}$ & H$\beta^*$ &
 H$\beta^{\dagger}$ & He~\textsc{i}$^*$ & He~\textsc{i}$^{\dagger}$ & He~\textsc{i} & He~\textsc{i} \\
Frame ID & 4387 & 4471 & 4686      & 4686              &         &   &
 4921 & 4921 & 5015 & 5875  \\ \hline
37106 & $-$88.6 &   $-$4.7 & $-$732.8 &  426.0 & $-$739.8 &  633.6 & $-$728.2 &
 675.7 &   20.7 &   33.3 \\
37107 & 118.7 &  105.9 & $-$507.7 &  509.4 & $-$578.5 &  647.8 & $-$572.6 &
 724.3 &  116.7 &   99.6 \\
37109 & $-$63.5 &  335.0 &        &        & $-$501.9 &        &        &
     &  264.4 &  381.9 \\
37110 & 142.7 &  326.7 & $-$650.8 &  981.4 & $-$686.9 &        & $-$624.5 &
 973.8 &   88.6 &  345.3 \\
37111 &  17.6 &   78.8 &        &  478.4 & $-$695.7 &        & $-$659.5 &
     &   82.4 &   46.0 \\
37112 & $-$44.7 &  $-$35.1 &        &        & $-$707.6 &        & $-$698.4 &
     &   54.8 &  $-$36.8 \\
37114 & $-$27.1 &   19.5 & $-$707.3 &  344.3 & $-$747.5 &  637.7 & $-$749.6 &
 704.3 &   39.3 &   20.3 \\
37115 &  45.4 &   67.1 & $-$713.6 & 1062.3 & $-$687.1 &  638.1 & $-$635.7 &
 713.1 &   81.9 &   87.4 \\
37116 & 209.6 &  189.9 & $-$470.2 &  675.3 & $-$533.3 &  817.1 & $-$512.9 &
 789.6 &  185.4 &  205.5 \\
37117 & 291.4 &  296.9 &        &        & $-$594.6 &  263.3 & $-$644.6 &
 1131.3 &  273.0 &  293.7 \\
37118 & 157.8 &  169.8 & $-$590.1 &  877.8 & $-$654.1 &  263.3 & $-$607.9 &
 956.5 &   82.2 &  321.2 \\
37121 & $-$57.5 & $-$107.1 &        &        & $-$763.0 &  645.1 & $-$726.7 &
 861.0 &    6.8 &  $-$69.5 \\
37122 & $-$20.9 &    6.2 & $-$784.8 &  584.1 & $-$789.6 &  607.6 & $-$751.5 &
 682.9 &   41.8 &   30.5 \\ \hline
\\
\end{tabular}

\begin{tabular}{crrrr}
\hline\hline
August 6 & H$\alpha^*$ & H$\alpha^{\dagger}$ & He~\textsc{i}$^*$ & He~\textsc{i}$^{\dagger}$ \\
Frame ID &  & & 6678 & 6678 \\ \hline
37106 & $-$678.3 &  458.9 & $-$660.5 &  633.5 \\
37107 & $-$565.6 &  480.7 & $-$581.1 &  628.2 \\
37109 &        &        & $-$450.9 &  743.1 \\
37110 &        &        & $-$631.1 &  766.8 \\
37111 & $-$625.5 &  420.0 & $-$650.3 &  549.7 \\
37112 & $-$615.4 &  269.0 & $-$670.2 &  614.9 \\
37114 & $-$673.6 &  460.3 & $-$687.1 &  624.8 \\
37115 & $-$636.6 &  481.5 & $-$661.8 &  611.7 \\
37116 & $-$528.1 &  523.1 & $-$498.9 &  632.7 \\
37117 & $-$509.2 &        &        &        \\
37118 & $-$621.5 &        & $-$601.1 &        \\
37121 & $-$652.1 &  305.3 & $-$697.2 &  618.5 \\
37122 & $-$717.7 &  463.6 & $-$704.8 &  586.4 \\ \hline
\end{tabular}
\end{center}
\end{table*}

\subsection{Radial Velocity Variations and Line Forming Regions}
\label{sec:RVs}

During the previous 1978 outburst, \citet{gil80wzsgeSH} found that
the radial velocities of the emission components of H$\alpha$ stayed
constant within $\pm$15 km~s$^{-1}$ and the peak separation of this line
($\sim$440 km~s$^{-1}$) did not vary for over one orbital period in the
spectra obtained in the 7th night of the outburst.  This separation did
not change from the 5th night to the 8th night.

\begin{figure*}
  \begin{center}
    \FigureFile(168mm,230mm){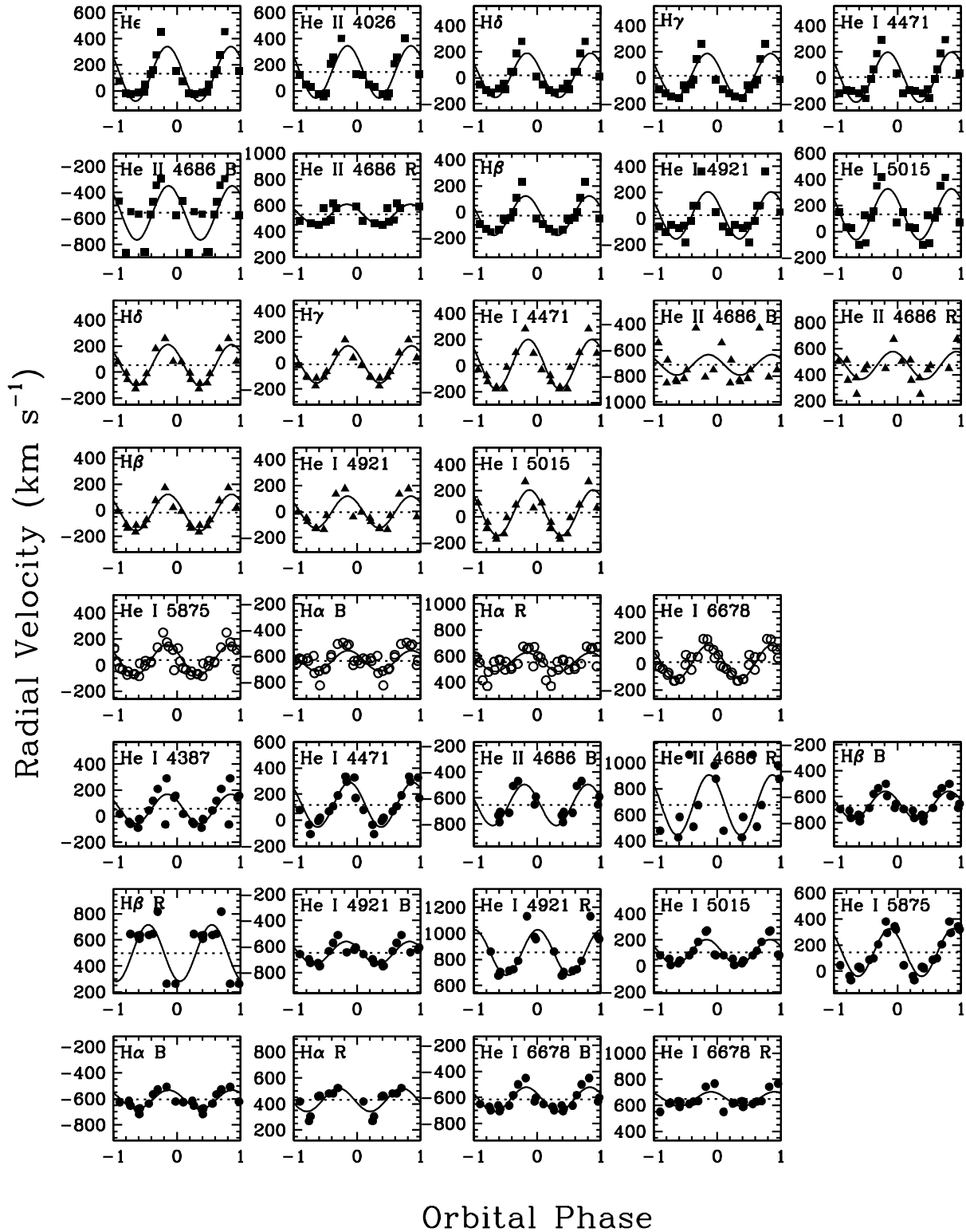}
  \end{center}
  \caption{Radial velocities of emission/absorption lines of July 30 and
 31, and August 1 and 6.  The marks represents the data in table
 \ref{tab:rv}, and the sine curves (table \ref{tab:fit1}) obtained by
 fitting the RVs are drawn in each figure.  The figures in the top 2
 rows (filled square), the third and fourth rows (filled triangle), the
 fifth row (open circle), and the bottom 3 rows (filled circle) displays
 RVs measured on 2001 July 30 and 31, and August 1 and 6, respectively.
  }
  \label{fig:rv}
\end{figure*}

Our observations, however, do not affirm the stationary H$\alpha$
emission.  The H$\alpha$ peak separation indicated an increase trend
(table \ref{tab:peak}), and the radial velocity changed with a large
amplitude.  Table \ref{tab:rv} summarizes the radial velocities of each
emission/absorption lines measured by fitting a Gaussian function during
the periods II and III.  Table \ref{tab:fit1} lists the parameters of
the semi-amplitude ($K$), the red-to-violet crossing time ($T'_0$), the
corresponding phase offset ($\phi_0$) and the systemic velocity $\gamma$
obtained by sinusoid fitting using the known orbital period.  The radial
velocities in table \ref{tab:rv} and the sinusoids with parameters in
table \ref{tab:fit1} are shown in figure \ref{fig:rv}.

Most of the phase offsets are within the range of $\phi_0=0.08-0.13$,
with exceptions being those of the H$\beta$ red peak, the red peak of He
\textsc{i} 4921, the H$\alpha$ red peak on August 6.  This phase shift
of $\phi_0\sim0.11$ may be against a hypothesis that the lines originate
from the accretion disk.  A shift of an almost same value was, however,
also observed in quiescence, which was interpreted to be due to an
effect by the hotspot contribution \citep{mas00wzsge}.  In the present
case, the spiral structures in the disk possibly have a stronger effect
(see \cite{bab02wzsgeletter,kuu02wzsge}).  Then, the possibility is not
rejected that the lines were formed on the accretion disk.

The reason for the unusual phase offsets of the red peaks of H$\beta$,
He \textsc{i} 4921, H$\alpha$ is left as an open problem.  Confirmation
of this phenomenon is desired during future outbursts.

The semi-amplitudes are larger than those in quiescence, except for
$\sim$75 km s$^{-1}$ of H$\alpha$ consistent with 68($\pm$3) km
s$^{-1}$ in quiescence \citep{mas00wzsge}, which probably suffer from
contamination of the spiral arms, too.

The absorption line of Na~\textsc{i} D definitely existed in our first
spectrum, though it is not clear in figure \ref{fig:firstspectra}
because of the contamination of He~\textsc{i} 5875.  As previously
mentioned, this absorption line was found in IY UMa \citep{wu01iyuma}
and EG Cnc \citep{pat98egcnc} in outburst.  \citet{pat98egcnc}
interpreted that an extensive cool region in the disk is an origin of
Na~\textsc{i} absorption, based on the broad FWZI.  Although the
contamination of the He~\textsc{i} absorption obstructs a measurement
of the FWZI of Na~\textsc{i} D in our spectra, we can estimate the
relative radial velocity using the profile from the bottom of the
absorption to the red wing.  The estimates in this way set limits on the
semi-amplitude of the radial velocity variation to be smaller than 25
km~s$^{-1}$, which is smaller than that of the white dwarf (see
\cite{ste01wzsgesecondary}).  Thus, the origin of the absorption is
suspected to be a rather stationary region, such as the center of mass
of this system, or the part of the disk extended over the secondary star
(see section \ref{sec:Rmax}).  It deserves attention that the absorption
depth reaches 0.08 in the normalized intensity (figure \ref{fig:IIsp}).

\begin{table}
\caption{Fitted semi-amplitudes (km s$^{-1}$), $T_0$
 ($BDJD-2452120$), systemic velocities (km s$^{-1}$) of lines on 2001
 July 30 and 31, and August 1 and 6.  The marks, $^*$ and $^{\dagger}$
 represent the blue component and the red component of the double peaks,
 respectively.}\label{tab:fit1}
\begin{center}
\begin{tabular}{lrrrr}
\hline\hline
July 30 \\
Line  & \multicolumn{1}{c}{K} & \multicolumn{1}{c}{$T_0$} &
 \multicolumn{1}{c}{$\phi_0$} & \multicolumn{1}{c}{$\gamma$} \\ \hline
H$\epsilon$                     & 210(31) & 1.442(1) & 0.10(2) & 131(20) \\
He~\textsc{ii}~4026             & 200(43) & 1.442(2) & 0.10(4) & 145(27) \\
H$\delta$                       & 170(29) & 1.441(1) & 0.09(2) & 19(19)  \\
H$\gamma$                       & 170(33) & 1.441(1) & 0.09(2) & 17(21)  \\
He~\textsc{i}~4471              & 192(38) & 1.442(1) & 0.10(2) & 5(25)   \\
He~\textsc{ii}~4686$^*$         & 208(85) & 1.443(3) & 0.12(5) & $-$557(54) \\
He~\textsc{ii}~4686$^{\dagger}$ &  76(32) & 1.441(3) & 0.09(5) & 533(20) \\
H$\beta$                        & 151(29) & 1.440(2) & 0.07(5) & $-$28(19) \\
He~\textsc{i}~4921              & 181(49) & 1.442(2) & 0.10(4) & 24(31)  \\
He~\textsc{i}~5015              & 195(55) & 1.442(2) & 0.10(4) & 133(35) \\
\hline
\\
\hline\hline
July 31 \\
Line  & \multicolumn{1}{c}{K} & \multicolumn{1}{c}{$T_0$} &
 \multicolumn{1}{c}{$\phi_0$} & \multicolumn{1}{c}{$\gamma$} \\ \hline
H$\delta$                       & 157(23) & 2.463(1) & 0.11(2) & 55(17) \\
H$\gamma$                       & 143(15) & 2.463(1) & 0.11(2) & $-$11(11) \\
He~\textsc{i}~4471              & 191(26) & 2.463(1) & 0.11(2) & 10(19) \\
He~\textsc{ii}~4686$^*$         &  78(66) & 2.463(8) & 0.11(14) & $-$714(48) \\
He~\textsc{ii}~4686$^{\dagger}$ & 106(44) & 2.467(5) & 0.18(9) & 473(36) \\
H$\beta$                        & 140(16) & 2.463(1) & 0.11(2) & $-$18(12) \\
He~\textsc{i}~4921              & 122(30) & 2.462(2) & 0.10(2) & $-$4(22) \\
He~\textsc{i}~5015              & 175(27) & 2.464(1) & 0.13(2) & 30(20) \\
\hline
\\
\hline\hline
August 1 \\
Line  & \multicolumn{1}{c}{K} & \multicolumn{1}{c}{$T_0$} &
 \multicolumn{1}{c}{$\phi_0$} & \multicolumn{1}{c}{$\gamma$} \\ \hline
He~\textsc{i}~5875  & 117(17) & 2.575(1) & 0.09(2) & 39(12) \\
H$\alpha^*$         &  78(20) & 2.576(2) & 0.11(4) & $-$636(14) \\
H$\alpha^{\dagger}$ &  72(21) & 2.575(2) & 0.09(4) & 553(15) \\
He~\textsc{i}~6678  & 130(14) & 2.576(1) & 0.11(2) & 14(10) \\
\hline
\\
\hline\hline
August 6 \\
Line  & \multicolumn{1}{c}{K} & \multicolumn{1}{c}{$T_0$} &
 \multicolumn{1}{c}{$\phi_0$} & \multicolumn{1}{c}{$\gamma$} \\ \hline
He~\textsc{i}~4387              & 113(36) & 8.471(3) & 0.10(5) & 56(25) \\
He~\textsc{i}~4471              & 171(22) & 8.473(1) & 0.13(2) & 118(15) \\
He~\textsc{ii}~4686$^*$         & 158(40) & 8.468(1) & 0.05(2) & $-$654(26) \\
He~\textsc{ii}~4686$^{\dagger}$ & 259(126)& 8.472(5) & 0.12(9) & 676(81) \\
H$\beta^*$                      & 107(19) & 8.469(2) & 0.06(4) & $-$666(13) \\
H$\beta^{\dagger}$              & 217(86) & 8.454(3) & $-$0.20(5) & 499(50) \\
He~\textsc{i}~4921$^*$          &  88(19) & 8.470(2) & 0.08(4) & $-$650(13) \\
He~\textsc{i}~4921$^{\dagger}$  & 174(42) & 8.480(2) & 0.26(4) & 854(25) \\
He~\textsc{i}~5015              &  98(19) & 8.470(2) & 0.08(4) & 105(13) \\
He~\textsc{i}~5875              & 182(27) & 8.473(2) & 0.13(4) & 143(18) \\
H$\alpha^*$                     &  73(18) & 8.472(2) & 0.12(4) & $-$606(12) \\
H$\alpha^{\dagger}$             &  89(33) & 8.463(5) & $-$0.04(9) & 432(27) \\
He~\textsc{i}~6678$^*$          &  94(19) & 8.470(2) & 0.08(4) & $-$615(13) \\
He~\textsc{i}~6678$^{\dagger}$  &  55(27) & 8.474(5) & 0.15(9) & 649(17) \\
\hline
\end{tabular}
\end{center}
\end{table}

\subsection{Emission lines and the outburst mechanism}\label{sec:emission}

The first spectra (ID: 10550) of our data was obtained at BDJD
2452114.48 (table \ref{tab:log}), which is several hours before the
outburst maximum.  He~\textsc{ii} 4686 was a doubly-peaked emission line
with a peak separation of $\sim$660 km s$^{-1}$ in this spectrum obtained
at $\phi=0.255$.  The spectrum (ID: 10557) obtained one hour after, but
still before the maximum, showed H$\alpha$ emission line with double
peaks of a $\sim$780 km s$^{-1}$ separation.  These small separations
mean that the accretion disk was greatly extended before the outburst
maximum (see the next section).  This qualitatively agrees with the
prediction on the outside-in type outburst due to the disk instability
(see e.g. \cite{ich93SHmasstransferburst}).

The spectra obtained by \citet{bab02wzsgeletter} by 5.5 hours before our
first observation contain only absorption lines of the Balmer series
including H$\alpha$ and He~\textsc{i} and no hint of emission lines of
highly ionized species: He~\textsc{ii} 4686,
C~\textsc{iii}/N~\textsc{iii}.  It should be a natural scenario that the
innermost part of the accretion disk and the white dwarf was fully
heated up by the accreted mass, and irradiation by this region formed a
temperature-inversion layer (chromosphere) on the disk during this
interval of 5 hours, then the chromosphere produced these
high-excitation emission lines.

The C~\textsc{iv}/N~\textsc{iv} emission lines, which was observed on
the 1st and 3rd days of this outburst, do not bear a clearly
doubly-peaked shape.  On JD 2452114, the FWZI of this blend line is very
broad, 4000--5000 km s$^{-1}$, which is comparable of $\sim$5600
km~s$^{-1}$ of the combined line of the C~\textsc{iii}/N~\textsc{iii}
blend and He~\textsc{ii} 4686.  However, the FWZI of
C~\textsc{iv}/N~\textsc{iv} is much broader than that of H$\alpha$ (2200
km~s$^{-1}$), even taking into account the separation of the blended
lines.  These two facts (the non-doubly-peaked profile and the broad
FWZI) imply that the C~\textsc{iv} and N~\textsc{iv} lines originated in
the boundary layer between the accretion disk and the white-dwarf
surface, or in the very inner region of the accretion disk.

On the other hand, there still remains a possibility that the
C~\textsc{iv}/N~\textsc{iv} complex was formed in the chromosphere in
the accretion disk, like He~\textsc{ii} 4686.  Our instruments may not
have a resolution power to resolve their doubly-peaked profiles.  Higher
spectral-resolution (and time-resolved) spectroscopy from the initial
phase of the future outbursts are encouraged.

These evolution courses of the high-excitation lines are also reasonably
understandable within the outside-in outburst scheme of the disk
instability.  The C~\textsc{iv}/N~\textsc{iv} emission line weakened
from the 1st day to the 3rd day, and disappeared by the 5th day of the
outburst, indicating that the temperature of the very inner side of the
accretion disk started decreasing at the near-the-maximum phase of the
outburst.  However, He \textsc{ii} 4686 was present throughout the main
outburst.  It will be a challenge for the disk instability model to
reproduce the variation of the temperature and density structure in the
accretion disk which can produce the spectral evolution of these
high-excitation lines in the special case of WZ Sge.

\subsection{Maximum disk radius}\label{sec:Rmax}

The peak separation is generally regarded to be indicative of the
velocity at the outer edge of the accretion disk multiplied by $\sin i$,
where $i$ is the inclination angle.  The separation of H$\alpha$ was
740(40) km s$^{-1}$ on JD 2452114 (table \ref{tab:peak}), which was
about a half of 1450 km s$^{-1}$ in quiescence \citep{ski00wzsge}.  If
we assume the Keplerian circular velocity and the disk radius $r_{\rm
disk} = 0.3a$ ($a$: the binary separation) in quiescence
\citep{rob78wzsge}, then the radius at the maximum of the outburst is
suggested to exceed the binary separation ($r_{\rm disk} \sim 1.2a$). 
Such a circumstellar disk was first introduced by \citet{gil80wzsgeSH}
to explain the narrow peak separation ($\sim$440 km s$^{-1}$) of
stationary H$\alpha$ observed during the 1978 outburst of WZ Sge (see
also \cite{bro80wzsgespec}; \cite{fri81wzsge}).  The Na \textsc{i} D
absorption line may be formed in the circumbinary part of this extended
disk, though this hypothesis requires a significant contribution of that
part to the continuum light to produce the absorption depth of 0.08.

To create a circumstellar disk, the outer edge of the disk must extend
over the critical radii of the 2:1 resonance and the tidal truncation,
and the Roche lobe.  \citet{lei81wzsge} proposed a model to create the
external disk, or ring, that the radiation pressure at the outburst may
make gases pour out from the L2 point.  There has been, however, no
observational evidence for the mass flow via the L2 point.

Here, we should pay attention to the following two points: 1) the peak
separation on JD 2452114 was determined using only the two spectra
obtained at $\phi=0.847$ (ID: 10557) and $\phi=0.896$ (ID: 10558), and
2) we assumed the Keplerian velocity.  As for the first point, the peak
separation observed in a spectrum changes under the influence by the
variations of the emission profile, as mentioned above.  The
separation was, however, derived from the spectra obtained at different
orbital phases on JD 2452116 and 2452118 (table \ref{tab:log}), and
maintained small values ($<$900 km s$^{-1}$) by JD 2452118 (table
\ref{tab:peak}).  If we take the 900 km s$^{-1}$ separation, the
resultant radius of $r_{\rm disk} \sim 0.77a$ still exceeds the
theoretical maximum radius and the critical radius of the 2:1 resonance
\citep{osa03DNoutburst}.

The second point arises from a question whether the Keplerian velocity
law holds up to the outer edge of the outbursting disk, especially
around the outburst maximum.  For this problem, \citet{osa03DNoutburst}
used the value of 1,000 km s$^{-1}$ in the discussion of the maximum
disk radius, based on the Doppler maps of He \textsc{ii} and H$\alpha$
in \citet{bab02wzsgeletter}, although these maps seem to admit a choice
of a larger separation.

To discuss more about this topic, we need higher spectral- and
temporal-resolution spectroscopy around the outburst maximum, detailed
theoretical works and simulations on the behavior of the outbursting
accretion disk.

\subsection{Alternation between the early/genuine superhump}

According to \citet{pat02wzsge} and \citet{ish04wzsge}, the genuine
superhumps emerged on JD 2452126.  The period I and II are characterized
by the early superhumps, and the genuine superhump were observed during
the period III and later.

\begin{figure*}
 \begin{center}
  \FigureFile(168mm,230mm){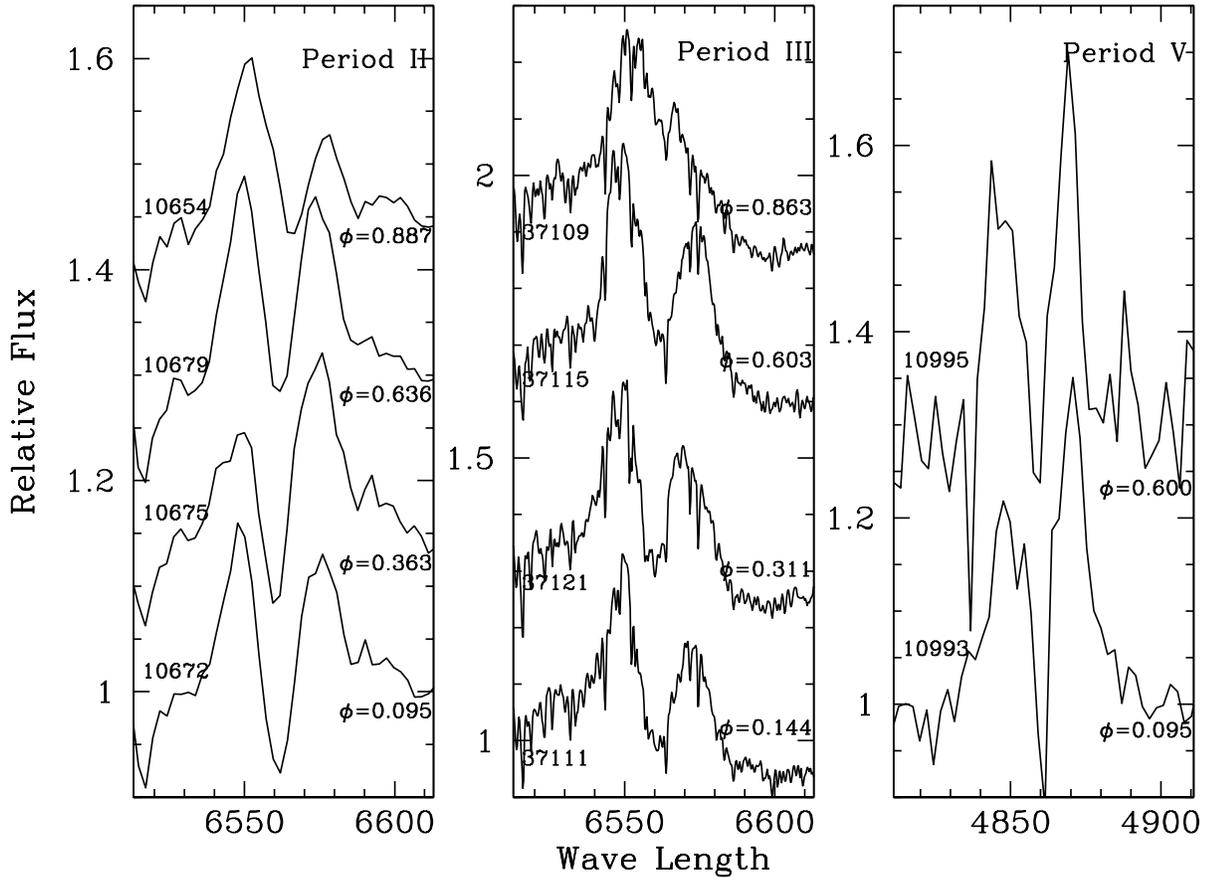}
 \end{center}
 \caption{Typical profile variations of H$\alpha$ in period II (left panel)
 and period III (mid panel) and H$\beta$ in period V (left panel) in an
 orbital period.  The stronger peak varies between the red peak and the
 blue peak in period II (early-superhump era), but one of the peaks
 dominates through the orbital phase during periods III and V
 (genuine-superhump era).
 }
 \label{fig:ha}
\end{figure*}

Figure \ref{fig:ha} displays the typical variations of the emission-line
profiles of H$\alpha$ in the periods II and III, and of H$\beta$ in the
period V in an orbital period.  The V/R ratio of the peaks of H$\alpha$
varied around 1 in period II when we saw early superhumps.  In contrast,
one of the peaks was stronger through the orbital phase in the periods
III and V, when genuine superhumps were prominent.  This clearly proves
the different nature between the early superhump and the genuine
superhump.

Interchange of the stronger peak with a beat period of the superhump
period and the orbital period has been interpreted to be evidence of the
precessing eccentric disk \citep{vog82zcha, hes92lateSH, aug94vyaqr,
wu01iyuma}.  The current plausible models for the early superhump
phenomenon are the tidal dissipation model \citep{osa02wzsgehump} and
the the tidal distortion model \citep{kat02wzsgeESH}.  The present data
indicate that the disk was not yet eccentric on JD 2452112 (section
\ref{sec:RVs}), the 10th day of the outburst and 4 days before the
genuine superhump emergence.  To test these model, it may be a key to
examine which model can more naturally allow a sudden change of the disk
structure which was observed photometrically
\citep{pat02wzsge,ish04wzsge} and spectroscopically (this work).

\subsection{Rebrightening phase}

The periods V and VI are around the third minimum and at the ninth peak
during the rebrightening phase, respectively.  All the Balmer lines and
He~\textsc{i} except for He~\textsc{i} 4471 are doubly-peaked emission
lines during the period V (figure \ref{fig:Vsp}), as in quiescence (see
e.g. \cite{dow81wzsge}), although the broad absorption component of the
white dwarf is not seen around any Balmer emissions.  During the period
VI (figure \ref{fig:VIsp}), the Balmer lines were basically in
absorption, although they seem to have weak emission components of
doubly-peaked shapes at the same time.  This change of the hydrogen and
helium absorption lines to emission lines with a change from a faint
state to a bright state means that propagation of the heating/cooling
wave gave rise to the state transition, as in the normal outburst of
dwarf novae (see \cite{osa01egcnc}).

The peak separations of the Balmer lines in table \ref{tab:peak} were
slightly narrower than those in quiescence, indicating that the
accretion disk had already shrunk to be $\sim$0.37$a$, close to that in
quiescence, and was well within the critical radius of the 3:1 resonance
(0.46$a$).  The eccentricity of the disk is, however, still maintained,
as indicated by the persistently strong red peak of H$\alpha$ in the
period V (see the previous section) and the asymmetric profile of
absorption lines in the period VI.

The high-excitation emission lines of He~\textsc{ii} 4686 and
C~\textsc{iii}/N~\textsc{iii}, of course also C~\textsc{iv} and
N~\textsc{iv}, are neither detected in the spectrum in the period V, nor
in the period VI, indicating the low temperature around the
boundary layer insufficient to strongly irradiate the outer disk.  The
Na~\textsc{i} absorption also can not be seen, while this line stayed
persistent during the main outburst.  These two differences of the
spectral feature between in the main outburst and in the rebrightening
phase may have the same origin, although the excitation potential of
these species is quite different.

\section{Summary and Conclusions}

Here we briefly summarize what we spectroscopically observed in optical
from the rising phase to the rebrightening phase of the 2001
outburst of WZ Sge, and what was revealed by detailed analyses of the
data and comparison of this outburst and the previous outbursts.

\begin{itemize}
 \item The variations of the outburst shape and the quiescence duration
       suggest an increasing trend of the mass transfer rate in these
       several tens of years.  The secondary star may still have
       magnetic activities, though the secondary has been supposed to be
       (close to) a degenerate star. (section 6.1)

 \item The radial velocities of the H and He emission/absorption lines
       measured in the main outburst had semi-amplitudes larger than
       those in quiescence, and the red-to-violet crossing times of
       these lines corresponded to $\phi \sim 0.11$.  These may be
       interpreted to suffer from an effect of the spiral-arm structures
       in the accretion disk.  (section 6.2)

 \item Na \textsc{i} D was found to be in strong absorption from the 1st
       day to, at least, 15th day of the outburst.  The semi-amplitude
       of the radial velocity variation of this line is limited to be
       within 25 km s$^{-1}$, which is smaller than that of the white
       dwarf.  A cool region might exist around the center of mass of
       this system, or the origin of this line may be the circumstellar
       part of the disk.  This absorption line was seen not only in WZ
       Sge stars (WZ Sge and EG Cnc), but also in an eclipsing SU UMa
       star, IY UMa in superoutburst.  (section 4, 5.1, 5.3, 6.2)

 \item Emergence of the high-excitation emission lines of He, C, and N,
       and the evolution from absorption to emission of H$\alpha$
       between $\sim$12 hours and $\sim$6 hours before the outburst
       maximum is a new evidence of the outside-in type outburst
       due to the disk instability.  The strong emission lines of
       C~\textsc{iv}/N~\textsc{iv} around 5800 \AA\ were first detected
       in the spectrum of dwarf novae, but disappeared by the 5th day of
       the outburst.  The emission lines of He~\textsc{ii} 4686,
       C~\textsc{iii}/N~\textsc{iii}, and H$\alpha$ is supposed to
       originate from the chromosphere of the accretion disk formed by
       irradiation. (section 4.1, 5.1, 6.3)

 \item The peak separations of H$\alpha$ and He~\textsc{ii} 4686 were
       about 700 km~s$^{-1}$, which is a half of that of H$\alpha$ in
       quiescence, at the very early phase of the outburst.  This
       implies that the accretion disk extended to have a circumstellar
       part. However, the phase coverage of our data are not sufficient
       for deriving the representative peak separation, and it is not
       clear whether the peak separation is naively used for measurement
       of the disk radius, especially around the outburst maximum
       (section 4.1, 5.1, 6.4)

 \item During the period of genuine superhumps (the latter half of the
       main outburst and the rebrightening phase), one of the double
       peaks of the Balmer emission lines dominated throughout one
       orbital phase, and the stronger peak interchanged in a time scale
       of days.  These represent the eccentricity of the accretion disk,
       which agrees with the tidal instability model of the superhump.
       Such characteristics were not found in the first half of the main
       outburst, at least four days before the emergence of the genuine
       superhump (8 days after the outburst maximum), when early
       superhumps were dominant periodic signals.  The eccentricity of
       the disk abruptly grew during this 4-day interval.  (section 6.5)

 \item During the rebrightening phase, the spectra showed absorption
       lines of H and He~\textsc{i} at the maximum and emission lines of
       the same species at the bottom, which means that the state
       transition of the disk was due to propagation of the
       heating/cooling waves, as in normal outbursts of usual dwarf
       novae.  There was no hints of the high-excitation lines and the
       Na~\textsc{i} absorption during this phase.  (section 4.5, 4.6,
       5.1, 6.6)

 \item Comparison of our spectra and those in the previous outbursts of
       WZ Sge indicates that the spectral feature and its evolution are
       different in each outburst.  We need further observations in
       future outbursts to reveal the whole nature of the king of dwarf
       novae.  (section 4, 5.2)

\end{itemize}

Finally, we would like to call the readers' attention for the emission
lines of He~\textsc{ii}.  In some spectra, e.g. figures 2 and 3,
He~\textsc{ii} are seen as emission lines, whereas He~\textsc{i} and
H~\textsc{i} are absorptions. If we use the intensity ratio of the
emission lines of He~\textsc{ii} 4686/H$\beta$ of these spectra to
estimate the helium abundance, we will have an abundance of infinity,
because there is no emission of H$\beta$.

Recently, \citet{iij02uscospec} estimated the helium abundance of a
recurrent nova U Sco using the intensity ratios of
He~\textsc{i}/H~\textsc{i}, and obtained as N(He)/N(H) = 0.16 by number,
which is normal among cataclysmic variables. On the other hand,
extremely high helium abundances of the same object have been derived
using the intensity ratios of He~\textsc{ii}/H~\textsc{i}, e.g. 0.4
\citep{anu00usco}, 2.0 \citep{bar81usco}, and 4.5 \citep{eva01uscoIR}.
The mystery of the helium abundance of U Sco might be due to a similar
effect. It seems to be rather risky to estimate the helium abundances of
cataclysmic variables using only the intensity ratios of
He~\textsc{ii}/H~\textsc{i}.

\vskip 3mm

The authors are thankful to amateur observers for reporting the
detection of this outburst their and continuous observations to VSNET.
Those reports enabled us to observe the most enigmatic dwarf nova WZ Sge
in a very precious phase before the maximum of a quite infrequent
outburst and to easily relate the spectral feature with the state of the
accretion disk.  Sincere thanks are also to Warren Skidmore and Yoji
Osaki for their valuable comments and discussions.

\end{document}